\documentclass[12pt]{article}
\usepackage{amsmath}
\usepackage{graphicx,psfrag,epsf}
\usepackage{enumerate}
\usepackage{natbib}
\usepackage{url} 

\newcommand{\blind}{0}

\usepackage[table]{xcolor}
\usepackage{amsthm}
\usepackage{mathtools}
\usepackage{collcell}
\usepackage{dsfont}
\usepackage{rotating}
\usepackage{wrapfig}
\usepackage[utf8]{inputenc}
\usepackage{tikz}
\usepackage{lettrine}
\usepackage{dblfloatfix}
\usepackage[english]{babel} 
\usepackage{amsfonts}
\usepackage{subfig}
\graphicspath{ {fig/} }
\usepackage[hyperfootnotes=false]{hyperref}

\usepackage{lipsum}

\usepackage{amsmath}

\DeclareMathOperator*{\argmin}{arg\,min}

\begin{document}

	\def\spacingset#1{\renewcommand{\baselinestretch}%
		{#1}\small\normalsize} \spacingset{1}

	
	\if0\blind
	{
		\title{\bf Estimation and simulation of the transaction arrival process in intraday electricity markets}
		\author{Michał Narajewski\hspace{.2cm}\\
			University of Duisburg-Essen\\
			and \\
			Florian Ziel \\
			University of Duisburg-Essen}
		\maketitle
	} \fi
	
	\if1\blind
	{
		\bigskip
		\bigskip
		\bigskip
		\begin{center}
			{\LARGE\bf Title}
		\end{center}
		\medskip
	} \fi
	
	\bigskip
	\begin{abstract}
		We examine the novel problem of the estimation of transaction arrival processes in the intraday electricity markets. 
		We model the inter-arrivals using multiple time-varying parametric densities based on the generalized F distribution estimated by maximum likelihood.
		We analyse both the in-sample characteristics and the probabilistic forecasting performance.  
		In a rolling window forecasting study, we simulate many trajectories to evaluate the forecasts and gain significant insights into the model fit. The prediction accuracy is evaluated by a functional version of the MAE (mean absolute error), RMSE (root mean squared error)
		and CRPS (continuous ranked probability score)
		for the simulated count processes.
		This paper fills the gap in the literature regarding the intensity estimation of transaction arrivals and is a major contribution to the topic, yet leaves much of the field for further development. 
		The study presented in this paper is conducted based on the German Intraday Continuous electricity market data, but this method can be easily applied to any other continuous intraday electricity market. For the German market, a specific generalized gamma distribution setup explains the overall behaviour significantly best, especially as the tail behaviour of the process is well covered.
	\end{abstract}

	\noindent%
	{\it Keywords: intraday market, point process, inter-arrival time, 
		trajectory simulation, transaction time, electricity market, probabilistic forecasting, density estimation
		} 
	\vfill

	\newpage
	\spacingset{1.2} 
	
	
	\section{Introduction}
	Consecutive growth of number and volume of transactions in the intraday electricity markets is observed since the introduction of these markets. This results in a higher concern of the researchers regarding the intraday electricity markets. \citet{uniejewski2018understanding} and \citet{narajewski2018econometric} consider a very short-term point electricity price forecasting (EPF) of the $\text{ID}_3$-Price index in the German Intraday Continuous market.  \citet{andrade2017probabilistic} and \citet{monteiro2016short} conducted research regarding the electricity price forecasting in the Iberian intraday electricity market. They performed a probabilistic electricity price forecasting and a point EPF using artificial neural networks, respectively. \citet{ziel2017modeling} and \citet{kulakov2019impact} examine the impact of renewable energy forecasts, i.e. wind and solar energy, on the intraday electricity prices. The relationship between the fundamental regressors and the price formation in the intraday markets is studied by many other scientists, e.g. \citet{pape2016fundamentals} or \citet{gonzalez2015impact}.

\begin{figure}[t!]
	\centering
	\includegraphics[width = 1\linewidth]{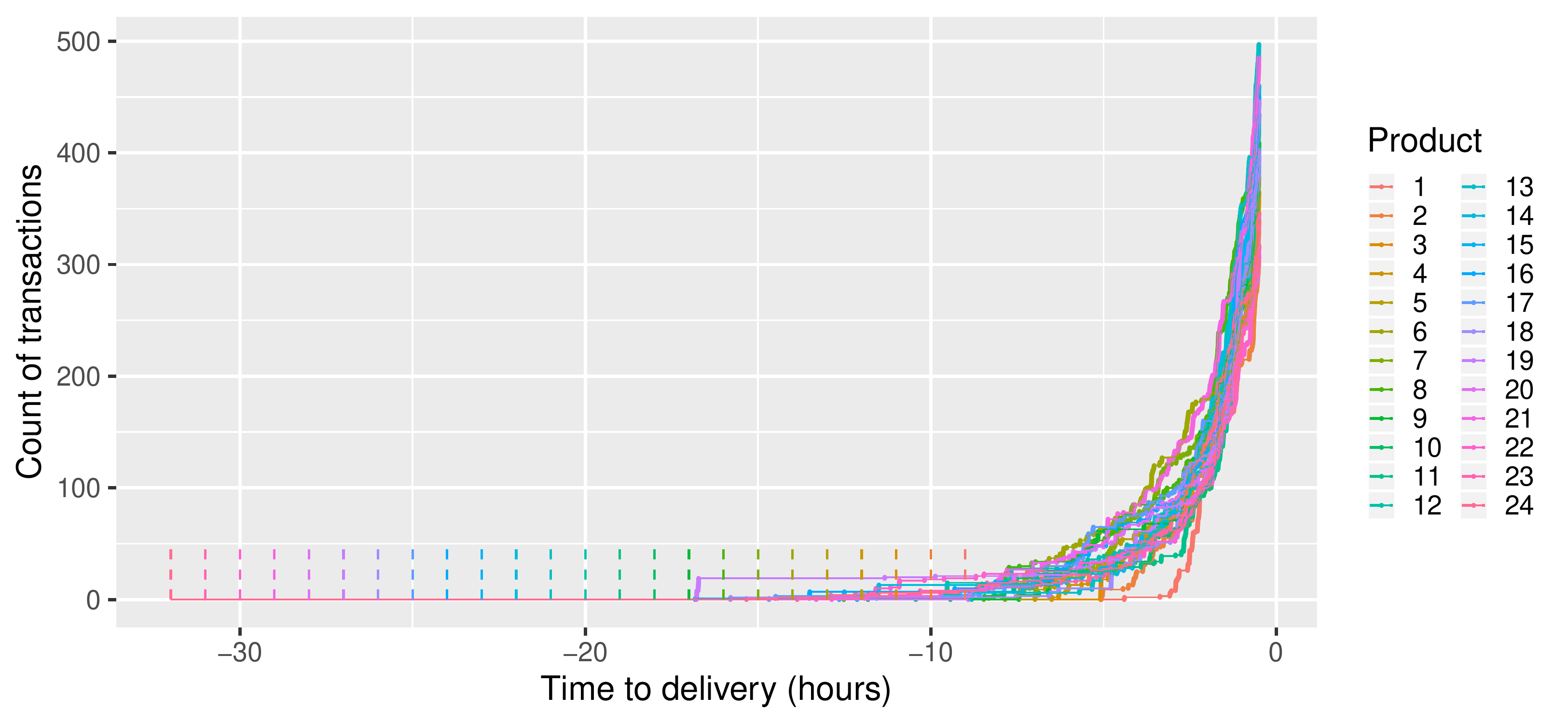}
	\caption{Trajectories of $S=24$ counting processes corresponding to transaction times for day $d = 03.09.2017$ in the hourly German Intraday Continuous market. The dashed lines indicate the beginning of trading period $b(d,s) = -8-s$ of product $s\in\{1,\ldots,S\}$ with respect to the day of delivery $d$, the end of delivery $e(d,s)=0.5 $. }
	\label{fig:first_trajectory}
		\end{figure}

	Due to the continuity of the intraday markets, an important aspect is the bidding behaviour of the market participants. This problem has been already examined by, among others, \citet{Kiesel2017} and \citet{aid2016optimal}. In the following paper, we take a closer look at the transaction arrivals in the intraday electricity market. Figure \ref{fig:first_trajectory} shows the trajectories of the counting processes that correspond to the transaction time arrivals. In the exercise, we assume that the transactions arrive in accordance with some time-dependent intensity function. Moreover, assuming the parametric probability distribution of the inter-arrival times, we can perform a maximum likelihood estimation of the parameters and then a meaningful forecasting study. In the exercise, we use a rolling window study, following the recommendations of \citet{diebold2015comparing}. 
	
	Our approach to the transactions in the intraday electricity markets is, to the best of our knowledge, an innovative one despite its simplicity.  \citet{von2017optimal} modelled the intensities of the buy/sell orders in a more complex manner, but they do not perform any forecasting study. 
	
			The outcome of the study is very satisfying despite the simplicity of utilized methods.
		The paper contributes to the literature by
		i) filling the gap in the literature regarding the intensity estimation of transaction arrivals and is a major contribution to the topic,
		
		ii) proposing a novel modelling approach for inter-arrival times using a time-varying generalized F-distribution to capture the underlying uncertainties
		
		iii) presenting a procedure to simulate trajectories from the estimated processes,
		
        iv) discussing functional evaluation criterion in forecasting studies for point processes,

		v) presenting a forecasting study for the German Intraday Continuous electricity market data.

		Note that the methodology can be easily applied to any other continuous intraday electricity market. For the German market, a specific generalized gamma distribution setup explains the overall behaviour significantly best, especially the tail behaviour of the process is well covered.

%

	The paper is structured as follows. In the next sections, we present the setting and the modelling details. Then, we explain the estimation and simulation methods. In the next section, we briefly describe the German Intraday Continuous market. In the sixth section, we introduce the forecasting study design, and we shortly present the exemplary dataset. Moreover, the evaluation measures are described. Next, the empirical forecasting results are presented. The paper is concluded with a discussion of the results and further development possibilities.

	\section{Setting}
In the majority of all continuous intraday markets there are $S$ products traded each day, e.g. $S=24$ in a market with hourly products.
For a certain day of delivery $d$ a product $s \in \{1,2, \dots, S\}$ is traded in the trading period $[b(d,s), e(d,s))$ prior to the beginning of the delivery.
Here $b(d,s)$ denotes the beginning of the trading and $e(d,s)$ the end of the trading. Both $b(d,s)$ and $e(d,s)$ potentially depend on the day of trading $d$ and the considered product $s$. However, in the majority of European intraday markets the end of trading does not depend on the product $s$ but the beginning of trading does. Furthermore, if not mentioned otherwise all times are measured in hours.

During the trading period $[b(d,s), e(d,s))$ we observe a series of $n(d,s)$ intraday 
transaction times $\mathbb{T}^{d,s} = (T^{d,s}_{1}, T^{d,s}_{2}, \dots T^{d,s}_{n(d,s)})$ satisfying 
$b(d,s)< T^{d,s}_{1}$,
$T^{d,s}_{i-1} < T^{d,s}_{i}$ for $i\in \{2,\ldots,n(d,s)\}$ and $T^{d,s}_{n(d,s)} < e(d,s)$.
An example of trajectories of corresponding counting processes is presented in Figure \ref{fig:first_trajectory} for $S=24$ products in the German Intraday Continuous market. 
As mentioned, the beginning of trading time differs for each product. For instance, the trading period in the German Intraday Continuous market for the hourly product $s=1$ with delivery starting at 00:00 is $[b(d,1),e(d,1)) = [-9, -0.5)$ and for the hourly product $s=24$ with delivery starting at 23:00 is $[b(d,24),e(d,24)) = [-32, -0.5)$.

Let us note that most of the transactions take place in the last hours of the trading period. The reason for this behaviour is the design of the intraday electricity market, i.e. its main purpose is to let the market participants react to the changes in production prediction. In the first hours of trading in the intraday market usually there is not much more information, when comparing to the day ahead market, but in the last hours before the delivery the difference is significant, and thus it is the most traded time period in this market. 
This pattern justifies the decision to parametrize the time in such a manner that the last hours of trading are indexed in the same way, disregarding the delivery time, as in Figure \ref{fig:first_trajectory}.

	
	\section{Modelling and estimation}
	In the purpose of estimating transaction arrivals, we consider the series of inter-arrival times $ (X^{d,s}_i) = (T^{d,s}_{i} - T^{d,s}_{i-1})$, where $i \in \{1,2,\dots, n(d,s)\}$, $n(d,s)$ is the number of transactions on day $d$ and product $s$ and $T_0^{d,s} = b(d,s)$ is the beginning of trading. 
	As pointed out in Figure \ref{fig:first_trajectory} only the latter hours of trading are of major interest for modelling. Hence,
	we focus on modelling only the part $[a(d,s), e(d,s))$ by choosing 
	$a(d,s)$ such that $b(d,s)<a(d,s)<e(d,s)$. In the example
	Figure \ref{fig:first_trajectory}, a reasonable choice for $a(d,s)$ could be e.g. $-8$, $-5$ or $-3$. Denote 
	$\mathbb{X}^{d,s} = (X^{d,s}_i)$ all inter-arrival times after $a(d,s)$, so they satisfy $ a(d,s) < T^{d,s}_{i}$. Further, let $l(d,s)$ be the smallest index such that $a(d,s) < T^{d,s}_{l(d,s)}$ holds.

	Now, we assume that the series of
	inter-arrival times $\mathbb{X}^{d,s}$ is independent and follows a probability distribution with a parametric density function $f_{\mathbb{X}^{d,s}}(x ; \theta)$. Therefore, knowing that the inter-arrival times are independent, we can perform the maximum likelihood estimation of the unknown vector of parameters $\theta$
	\begin{equation}
	\begin{aligned}
	\widehat{\theta} =  & \arg\max_\theta f_{\mathbb{X}^{d,s}}(x;\theta) = \arg\max_\theta \prod_{i=
	l(d,s)}^{n(d,s)} f_{\mathbb{X}^{d,s}}(x^{d,s}_i; \theta).
	\end{aligned}
	\end{equation}
	Naturally, to make the estimation less biased, we can estimate the parameters using more than one day of history of the transaction arrival times. Assuming the independence between them and that we estimate based on $D$ days of history, we get the following maximum likelihood estimator
	\begin{equation}
	\begin{aligned}
	\widehat{\theta} =  & \arg\max_\theta f_{(\mathbb{X}^{1,s}, \mathbb{X}^{2,s}, \dots, \mathbb{X}^{D,s})}(x;\theta) \\ = & \arg\max_\theta \prod_{d=1}^{D} f_{\mathbb{X}^{d,s}}(x;\theta) = \arg\max_\theta \prod_{d=1}^{D} \prod_{i=l(d,s)}^{n(d,s)} f_{\mathbb{X}^{d,s}}(x_i^{d,s};\theta).
	\end{aligned}
	\label{eq:maxlikel}
	\end{equation}
	The maximum likelihood problem stated in \eqref{eq:maxlikel} is solved using \verb+Rsolnp+ package in \verb+R+, which was implemented by \citet{rsolnp}, based on the algorithm of \citet{rsolnpphd}, which is the general non-linear augmented Lagrange multiplier method.  Since the likelihood function may contain local maxima, it is very important to set correctly the lower and upper bounds and the starting parameters. The algorithm should handle with no big problem up to 10-parametric optimization, so in purpose of our study it is satisfactory. Nevertheless, the choice of the maximum likelihood optimization tool is not crucial as we have a low dimensional problem.
	
	In the case of German Intraday Continuous market, we assume four distributions of the inter-arrival times: exponential, gamma, generalized gamma and generalized F. Each of the consecutive distributions is an extension of the previous one. The distributions are parametrized as follows:
	\begin{itemize}
		\item exponential distribution Exp($\lambda$) with rate parameter $\lambda > 0$,
		\item gamma distribution Gamma($\alpha, \beta$) with shape and rate parameters $\alpha > 0$ and $\beta >0$,
		\item generalized gamma distribution GenGam($\mu, \sigma, Q$) with location, scale and shape parameters $\mu \in \mathbb{R}, \sigma >0$ and $Q \in \mathbb{R}$,
		\item generalized F distribution GenF($\mu, \sigma, Q, P$) with location and scale parameters  $\mu \in \mathbb{R}$ and $\sigma >0$, and shape parameters $Q \in \mathbb{R}$ and $P \geq 0$.
	\end{itemize}
	The exponential and gamma distributions are well-know and thus do not need any special introduction.  The exponential distribution has the following density function
	\begin{equation}
		f(x;\lambda) = \lambda \exp\{-\lambda x\}
	\end{equation} 
	and the gamma distribution has the density function defined by
	\begin{equation}
		f(x;\alpha,\beta) = \frac{\beta^\alpha}{\Gamma(\alpha)}x^{\alpha-1}\exp\{-\beta x\}.
	\end{equation}
	Let us remind that the exponential distribution is a special case of the gamma distribution, i.e. if $X \sim \text{Exp}(\lambda)$, then $X \sim \text{Gamma}(\alpha = 1,\beta = \lambda)$. 
	
	The generalized gamma distribution is an extension of the gamma distribution by \citet{stacy1962generalization}, but in the study we use the parametrisation of \citet{prentice1974log}, which is stated above. If $\gamma \sim \text{Gamma}(Q^{-2}, 1)$ and $w = \log(Q^2 \gamma)/Q$, then $x = \exp(\mu +\sigma w)$ follows the generalized gamma distribution with probability density function
	\begin{equation}
		f(x;\mu,\sigma,Q) = \frac{|Q|(Q^{-2})^{Q^{-2}}}{\sigma x \Gamma(Q^{-2})} \exp\{Q^{-2}(Qw - \exp\{Qw\})\}.
	\end{equation}
	The relationship between the gamma and generalized gamma distributions parametrized in such a manner is as follows. If $X \sim \text{Gamma}(\alpha,\beta)$, then $X \sim \text{GenGam}(\mu = -\log(\beta/\alpha), \sigma = 1/\sqrt{\alpha}, Q = 1/\sqrt{\alpha})$. 
	
\begin{figure}[t!]
		\centering
		\includegraphics[width = 1\linewidth]{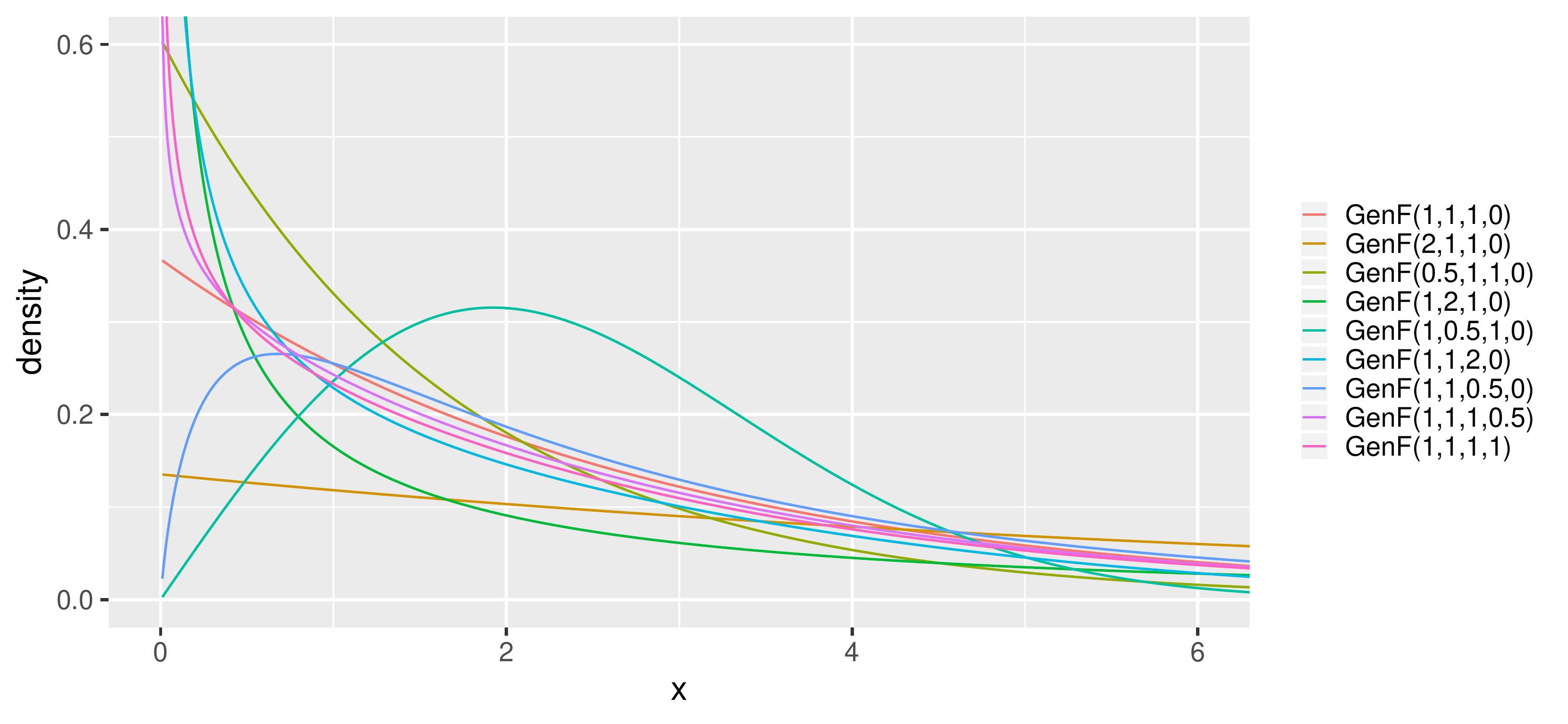}
		\caption{Illustration of the GenF distribution with different parameters}
		\label{fig:densities}
	\end{figure}

	The last and the most general distribution that we assume is the generalized F distribution described by \citet{prentice1975discrimination}. Define $s_1 = 2(Q^2+2P+Q\delta)^{-1}$ and $s_2 = 2(Q^2 +2P-Q\delta)^{-1}$, where $\delta = (Q^2 + 2P)^{1/2}$. If $w = (\log(x) - \mu)\delta/\sigma$, then the probability density function of $x$ is given by
	\begin{equation}
		f(x;\mu,\sigma,Q,P) = \frac{\delta (s_1/s_2)^{s_1}\exp\{s_1 w\}}{\sigma x (1+s_1 \exp(w)/s_2)^{(s_1+s_2)}B(s_1,s_2)},
	\end{equation}
	where $B(s_1,s_2)$ is the beta function.
	Let us note that if we possess a random variable $X \sim \text{GenGam}(\mu, \sigma, Q)$, then $X \sim \text{GenF}(\mu, \sigma, Q, P = 0)$. We see clearly that all the consecutive distributions are superior to the previous ones. Figure \ref{fig:densities} presents densities of exemplary GenF distributions.
	In the exercise, we use the \verb+flexsurv+ package in \verb+R+ by \citet{jackson2016flexsurv}, which contains the implementation of the generalized gamma and F distributions.

	
	We assume that the rate parameter $\lambda$, the rate and shape parameters $\beta$ and $\alpha$, and the location and scale parameters $\mu$ and $\sigma$ are some deterministic functions dependent on time $t$ and unknown vector of parameters $\theta$. Thus, we model them using the following functions:
	\begin{itemize}
		\item constant --- $f(t;\theta) = c$, where $\theta = c$,
		\item linear --- $f(t;\theta) = c + \beta_1 t$, where $\theta = (c,\beta_1)$,
		\item quadratic --- $f(t;\theta) = c+   \beta_1 t+\beta_2t^2$, where $\theta = (c, \beta_1, \beta_2)$,
		\item exponential --- $f(t;\theta) =  c + e^{\alpha_1+\alpha_2 t}$, where $\theta = (c, \alpha_1, \alpha_2)$.
	\end{itemize}
	In the estimation of the parameters of generalized gamma and F distributions, we make use of the relationship between gamma and generalized gamma distributions, i.e. $\mu(\theta,t) = -\log(\beta(\theta_2,t)/\alpha(\theta_1,t))$ and $\sigma(\theta,t) = 1/\sqrt{\alpha(\theta_1,t)}$. Moreover, we assume that $\alpha(\theta_1,t)$ cannot be more complex than $\beta(\theta_2, t)$ and that the $Q$ and $P$ parameters are constant over time to delivery. By complexity of a function, we mean the number of parameters. Using this criteria, quadratic and exponential functions are equally complex.
	
	The study consists in total of 37 models of the inter-arrival times process $\mathbb{X}^{d,s}$: 4 models with the assumption of the exponential distribution and 11 models per other considered distributions. We abbreviate them by \textbf{X.Y.Z}, where \textbf{X} stands for the distribution and \textbf{Y} and \textbf{Z} stand for the types of the $\beta(\theta_2, t)$ and $\alpha(\theta_1,t)$ functions, respectively. For instance, \textbf{Gamma.Lin.Const} stands for a model with assumed gamma distribution, linear $\beta(\theta_2, t)$ and constant $\alpha(\theta_1,t)$. Let us note that for the model with exponential distribution and $\lambda(\theta, t) = \theta$ the corresponding counting process is a homogeneous Poisson process and this model is our basic benchmark. Making the intensity function $\lambda(\theta, t)$ non-constant is the first extension of the benchmark and changing the distribution to the more general one is the further extension. Moreover, in this study for non-constant rate and shape functions we actually assume that the functions are constant in short intervals, e.g. for exponential distribution, $\lambda(\theta, T^{d,s}_i)$ is assumed to be the intensity on the time interval $[T^{d,s}_{i}, T^{d,s}_{i+1})$. This means that the corresponding counting process of a model with exponential distribution is a mixture of homogeneous Poisson processes. 
		\begin{figure}[t!]
		\centering
		\includegraphics[width = 1\linewidth]{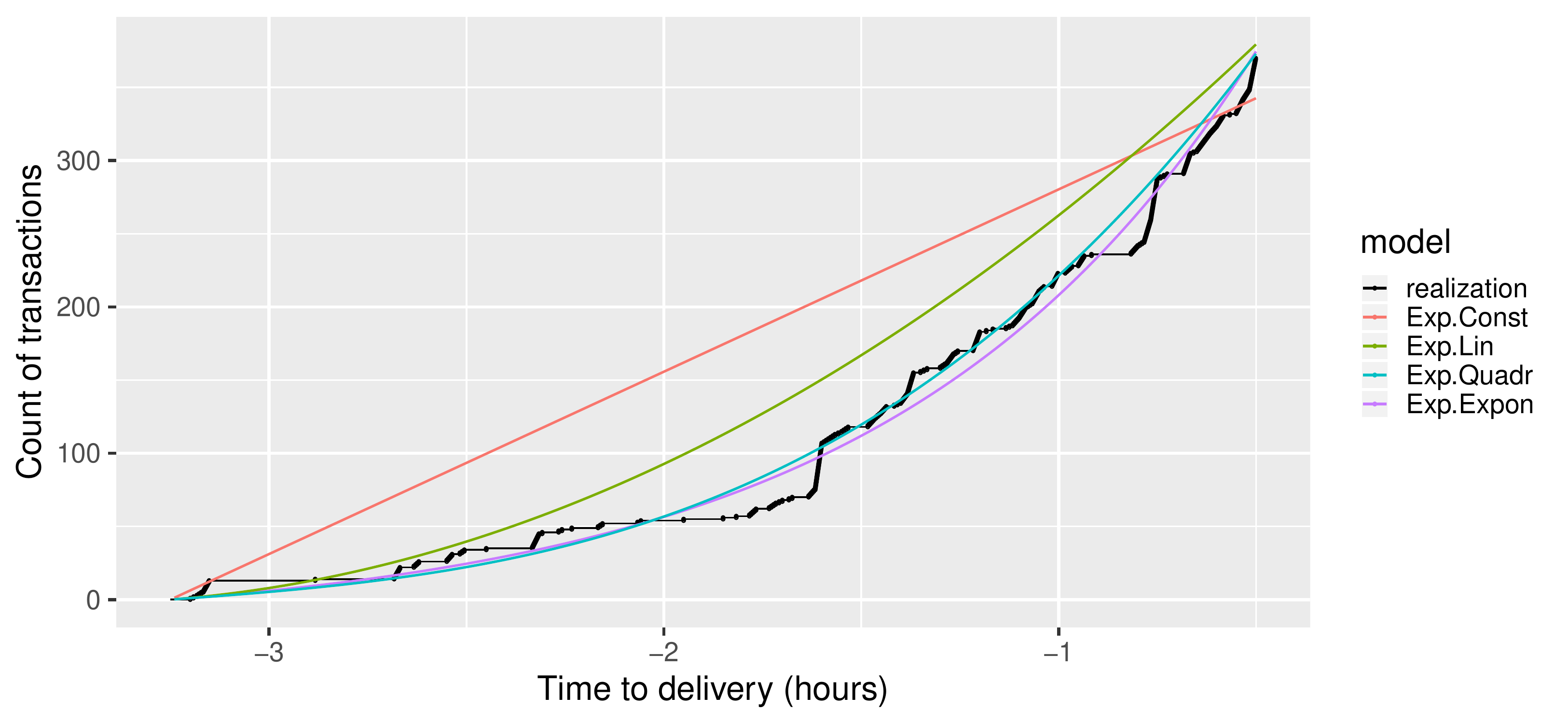}
		\caption{Observed trajectory of transaction counting process and fitted cumulative intensities of considered intensity functions of exponential distribution models for day $d = 03.09.2017$ and product $s=12$ in the hourly German Intraday Continuous market}
		\label{fig:cumulative_intensities}
	\end{figure}

	Let us note that the models contain not more than 8 parameters, so their estimation should not be a problem. Figure \ref{fig:cumulative_intensities} shows an example of fitting the aforementioned models assuming the exponential distribution to one day of data for one product. The figure presents the observed trajectory and estimated cumulative intensity functions $\Lambda(B) = \int_B \lambda(t)dt$. The time to delivery range is $[a(d,s),e(d,s)) =[-3.25, -0.5)$, because in this particular exercise we aim to forecast the transaction arrivals during the $\text{ID}_3$-Price period. This approach to the German Intraday Continuous was taken also by other researchers (\citealp{narajewski2018econometric}; \citealp{uniejewski2018understanding}).  Based on Figure \ref{fig:cumulative_intensities}, we may expect that considering the exponential distribution models, the models with quadratic and exponential intensity functions have similar performance in our problem.

	\section{Simulation}

	The trajectory simulation is relatively easy in our setting. Since we assume the distribution of the inter-arrival times $\mathbb{X}^{d,s} = (X^{d,s}_i)$, we can simply generate the inter-arrival times from the estimated distribution and calculate the next arrival time by adding the simulated  inter-arrival time to the time of forecasting. The only not that obvious part is the simulation of the first transaction arrival. That is to say, if we simulate at the time of the last observed transaction, there is no change, but if we fix the time of forecasting (e.g. like we do in the study), then we have to truncate the distribution for the first observation. This truncated distribution has the following density function
	\begin{equation}
		 f_{\mathbb{X}^{d,s}}\left(\theta, x_i^{d,s} | X_i^{d,s} > y\right) = \frac{g(x_i^{d,s})}{1- F_{\mathbb{X}^{d,s}}(y)},
	\end{equation}
	where $g(x) = f(x)$ for all $x > y$ and $g(x) = 0$ otherwise, and $F_{\mathbb{X}^{d,s}}(y)$ is the cumulative distribution function of $\mathbb{X}^{d,s}$. All the next transaction arrivals are simulated using the standard distribution with the density function $f_{\mathbb{X}^{d,s}}$ at the simulated times $\widehat{T}_{j}$, where $j = i + 1, i+2, \dots$ until the desired end of forecasting.
	
			\begin{figure}[b!]
		\centering
		\includegraphics[width = 1\linewidth]{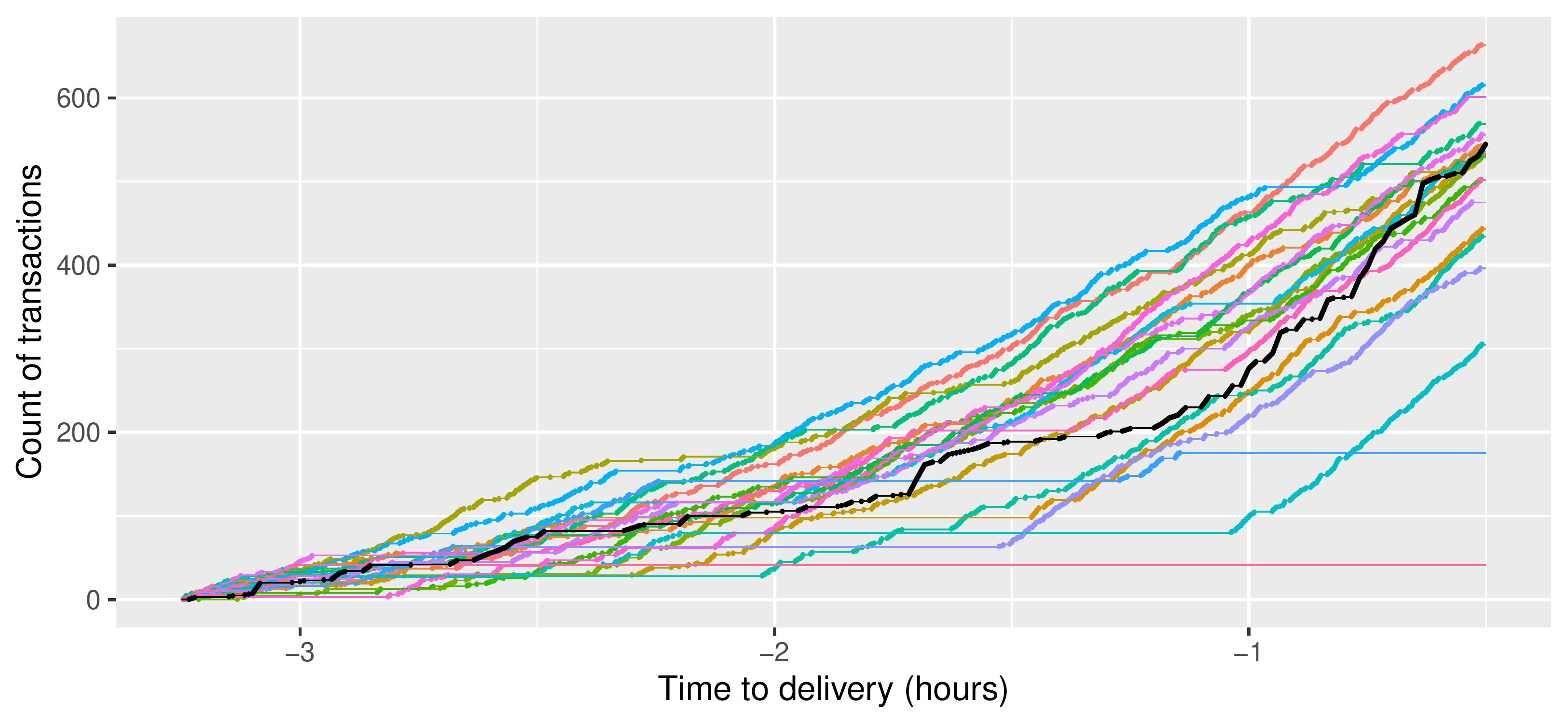}
		\caption{Observed (black) and simulated from the \textbf{GenGam.Quadr.Expon} (colourful) trajectories of transaction counting process for day $d = 01.10.2017$ and product $s=18$ in the hourly German Intraday Continuous market}
		\label{fig:example_trajectories}
	\end{figure}
	
	An example of simulation of 20 trajectories in the German Intraday Continuous market is shown in Figure \ref{fig:example_trajectories}. The trajectories are simulated as described above, based on the generalized gamma distribution with quadratic $\alpha(\theta_1,t)$ and exponential $\beta(\theta_2,t)$ functions, which was estimated using $D = 28$ days of data.

	\section{German Intraday Continuous market}
\begin{figure*}[!b]
		\begin{tikzpicture}[scale=1.07]
		\draw [->] [ultra thick] (0,0) -- (14.9,0);
		\draw [line width = 1] (1.5,1) -- (1.5, -0.5);
		\node [align = center, below, font = \scriptsize] at (1.5, -0.5) {$d-1$,\\ 12:00};
		\node [align = center, above, font = \scriptsize] at (1.5,1) {Day-Ahead\\Auction};
		\draw [line width = 1] (4,1) -- (4, -0.5);
		\node [align = center, below, font = \scriptsize] at (4, -0.5) {$d-1$,\\ 15:00};
		\node [align = center, above, font = \scriptsize] at (4,1) {Intraday\\Auction};
		\node [align = center, below right, font = \scriptsize] at (4,0) {Hourly Intraday Continuous};
		\draw [line width = 1] (5.05,1) -- (5.05, -0.5);
		\node [align = center, below, font = \scriptsize] at (5.05, -0.5) {$d-1$,\\ 16:00};
		\node [align = center, above right, font = \scriptsize] at (5.05,0) {Quarter-Hourly\\ Intraday Continuous};
		
		\draw [line width = 1] (10.5,1) -- (10.5, -0.5);
		\node [align = center, above, font = \scriptsize] at (10.5,1) {End of trading\\ on the market};
		\node [align = center, below, font = \scriptsize] at (10.5, -0.5) {$d$,\\ $s - 30$ min.};
		
		\draw [line width = 1] (12.5,1) -- (12.5, -0.5);
		\node [align = center, above, font = \scriptsize] at (12.5,1) {End of trading\\ within\\ control zones};
		\node [align = center, below, font = \scriptsize] at (12.5, -0.5) {$d$,\\ $s - 5$ min.};
		\draw [line width = 1] (14,1) -- (14, -0.5);
		\node [align = center, above, font = \scriptsize] at (14,1) {Delivery};
		\node [align = center, below, font = \scriptsize] at (14, -0.5) {$d$, $s$};
		\end{tikzpicture}
		\caption{The daily routine of the German electricity market. $d$ corresponds to the day of the delivery and $s$ corresponds to the hour of the delivery.}
		\label{fig:market}
	\end{figure*}
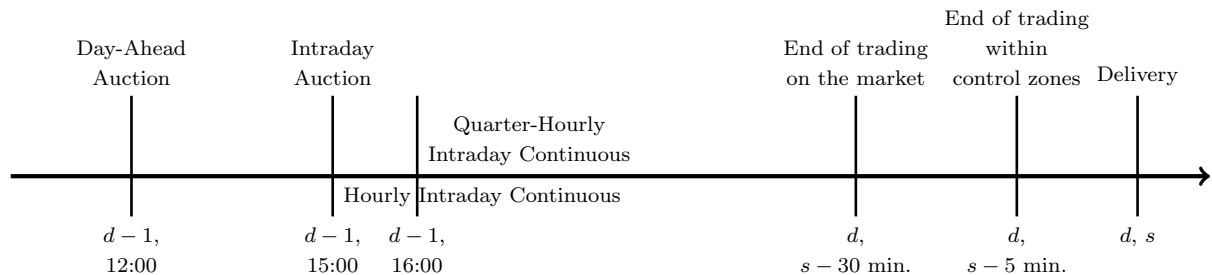

	As mentioned, in the exemplary study we consider the German Intraday Continuous market. Trading in this market starts every day at 15:00 for hourly and at 16:00 for quarter-hourly products of the following day. Market participants can trade electricity until 30 minutes before the delivery in the whole market and until 5 minutes before the delivery in respective control zones. Figure \ref{fig:market} presents briefly the German electricity spot market, for more details see \citet{Viehmann2017}.

	An important measure in the German Intraday Continuous market is the ID$_3$-Price index. The index is a volume-weighted price of all transactions taking place in the time interval between 3 hours and 30 minutes before the delivery and it is calculated separately for each intraday product. The importance of the ID$_3$-Price has been already noticed by the researchers and is a subject to modelling and forecasting by \citet{uniejewski2018understanding} and \citet{narajewski2018econometric}. In the latter paper, one can find a broader description and analysis of the ID$_3$-Price and the German Intraday Continuous market. In both papers, the authors performed a very short-term point EPF of the ID$_3$-Price. The outcome of \citet{narajewski2018econometric} is the efficiency of the market, i.e. the volume-weighted price of the transactions in the last 15 minutes before forecasting appears to be the best model for the ID$_3$-Price. In our study, we aim to forecast the time arrivals during the ID$_3$-Price time interval. Naturally, we leave ourselves some time for calculation and decision-making and therefore the considered time frame is $[-3.25, -0.5)$ which is exactly the same as in \citet{narajewski2018econometric}.
	
	In Europe the majority of intraday electricity markets features a similar structure to the German intraday market.
	This holds especially for all markets that 
	participate in the Cross-Border Intraday Project (XBID), see e.g. \citet{kath2019modeling}. It allows various 
	participating electricity markets (e.g. Germany, France, Spain) to bid across borders, if inter-connector capacity allows
	doing so. From the modelling perspective
	it might be relevant to note that all electricity markets which participate in XBID allow close their markets for each product the same amount of time before delivery, even though not always half an hour before delivery as in the German case. Thus, applying the modelling methodology to these markets should not be a problem.
	
	
	\section{Data, forecasting study and evaluation}
	In the following paper, as an example we perform a rolling window forecasting study based on the data from the German Intraday Continuous market. We consider a $D = 28$-day window size with the initial in-sample data from 03.09.2017 to 30.09.2017 and forecast the next day arrivals, starting on 01.10.2017. We forecast the arrivals between 3 h 15 min and 30 min before the delivery. Our out-of-sample study is of size $N = 365$, thus it spans the data range from 01.10.2017 to 30.09.2018. During each out-of-sample iteration $M = 1000$ trajectories are simulated. In the study, a multivariate approach is taken, which means that we create 24 separate models, each for every hourly product.
	
	The data that we utilize in the study was obtained from EEX Transparency, and it consists of information regarding: the date of the delivery, the product type, market area, volume of the traded energy, price in EUR/MWh, the transaction ID and the time of the transaction. In our case, the only relevant informations are the date of the delivery, the product type and of course the time of the transaction. A small inconvenience regarding the data is the fact that the transaction times have minute grid. This makes many transactions have identical time arrival even if they weren't made at the same time, e.g. 4 transactions with timestamp of 30.09.2017 16:01:00. We deal with the problem by distributing the transactions with the same timestamp $T$ uniformly in the time range $[T, T+ 1\text{min})$. Using the aforementioned timestamp as an example, the new timestamps are: 30.09.2017 16:01:00, 16:01:15, 16:01:30, 16:01:45.
	
	Due to the lack of literature regarding the intensity estimation in the intraday markets, we cannot use any literature benchmark models or literature evaluation measures. Thus, in the study we simply compare the results of simulation of all the considered models and as evaluation  measures we use the functional: bias (Bias), mean absolute error (MAE), root mean squared error (RMSE) and continuous ranked probability score (CRPS). We abbreviate the functional measures in a standard way, but the calculation is a little different. That is to say, let us denote by $N^{d,s}(t)$ the counting process of the true transactions on day $d$ for product $s$ and by $N_m^{d,s}(t)$ the $m$-th simulation of the counting process $N^{d,s}(t)$.
%
%
%

 Let $$\rho_{\eta,\tau,p}(z) = (\eta z^p+(1-\eta)|z|^p )|\tau-\mathds{1}(z<0)|$$ a 
 loss function with $\eta\in\{0,1\}$, $\tau\in(0,1)$, $p\geq 1$ and $\mathds{1}$ as indicator function. Then we define
\begin{equation}
 \widehat{N}^{d,s}_{\eta,\tau,p}(t)  = \argmin_{z} \sum_{m=1}^M \rho_{\eta,\tau,p}( N_m^{d,s}(t)-z) 
 \label{eq_argmin_est}
\end{equation}
the sample  $\rho_{\eta,\tau,p}$-estimate of the corresponding simulation sample $N_1^{d,s},\ldots, N_M^{d,s}$. The special case 
$(\eta,\tau,p)=(0,0.5,2)$ in  \eqref{eq_argmin_est} corresponds to 
ordinary least squares (OLS) and 
$(\eta,\tau,p)=(0,0.5,1)$ to median regression.
Thus, $\widehat{N}^{d,s}_{0,0.5,2}$
is the sample mean process, 
$\widehat{N}^{d,s}_{0,0.5,1}$ the sample median process and similarly $\widehat{N}^{d,s}_{0,\tau,1}$ the sample $\tau$-quantile process on which we focus especially.
Now, we define the evaluation criteria
\begin{equation}
\begin{aligned}
\text{EVAL}^{d,s}_{(\eta_1,\tau_1,p_1), (\eta_2,\tau_2,p_2)}([T_1, T_2)) = & \left(\int_{T_1}^{T_2} \rho_{\eta_1,\tau_1,p_1} ( N^{d,s}(t) - \widehat{N}^{d,s}_{\eta_2,\tau_2,p_2}(t) ) dt \right)^{\frac{1}{p}}\\ = &\left( \sum_{j=1}^{J} \rho_{\eta_1,\tau_1,p_1} ( N^{d,s}(t_j) - \widehat{N}^{d,s}_{\eta_2,\tau_2,p_2}(t_j)   )  \Delta t_j  \right)^{\frac{1}{p}}
\end{aligned}
\end{equation}
on a time range $[T_1, T_2)$.
$J$ is the length of a grid of the time range $[T_1, T_2)$ with $t_0 = T_1$ and $t_J = T_2$. The grid is defined by the jumps of both of the counting processes.
Let us note that the transition from the integral to the sum is possible, because the difference of counting processes is a simple function.
Moreover, we approximate the values of the evaluation measures by using a minute grid instead of the one defined by the jumps to reduce computational costs. 

Now, we define the special cases of the evaluation measures which lead
to the functional bias (Bias), the functional MAE, functional RMSE and functional pinball (PB) loss (or quantile loss) with respect to a probability $\tau$:
\begin{align}
\text{Bias}_s &=
\frac{1}{N}\sum_{d=1}^N 2 \text{EVAL}_{(1,0.5,1),(1,0.5,2)}^{d,s} \label{eq_crit1} \\
\text{MAE}_s &=
\frac{1}{N}\sum_{d=1}^N 2 \text{EVAL}_{(0,0.5,1),(0,0.5,1)}^{d,s} \label{eq_crit2} \\
\text{RMSE}_s &=
\frac{1}{N}\sum_{d=1}^N 2 \text{EVAL}_{(0,0.5,2),(0,.5,2)}^{d,s} \label{eq_crit3} \\
\text{PB}_{\tau,s} &=
\frac{1}{N}\sum_{d=1}^N \text{EVAL}_{(0,\tau,1),(0,\tau,1)}^{d,s} \label{eq_crit4}
\end{align}
We may observe that $2\text{PB}_{0.5,s} = \text{MAE}_s$.
Additionally, we use the pinball loss to approximate the functional continuous ranked probability score (CRPS) by
\begin{equation}
\text{CRPS}_{s} = \frac{1}{R}\sum_{\tau\in r}\text{PB}_{\tau,s}
\label{eq_crit5}
\end{equation}
for an equidistant grid of probabilities $r$ between 0 and 1 of size $R$, see e.g. \citet{nowotarski2018recent}. We consider the choice of $r = (0.01,0.02,\ldots,0.99)$ of size $R=99$.
	 In the purpose of comparing the models' forecasting performance, we calculate the functional Bias, MAE, RMSE and CRPS based on $M =1000$ trajectories in $N = 365$ out-of-sample iterations and for all $s\in\{1,\ldots,S\}$ with $S=24$ products. 
	For creating summaries, it may be useful to average across all products and define
	\begin{equation}
	\text{Crit} = \frac{1}{S}\sum_{s=1}^S \text{Crit}_s 
	\label{eq_crit}
	\end{equation}
	for the considered criteria $\text{Crit} \in \{\text{Bias}, \text{MAE}, \text{RMSE}, \text{PB}_{\tau}, \text{CRPS}\}$.

 \color{black}
 
	To draw significant conclusions on the outperformance of the forecasts of the considered models, we also calculate the \citet{diebold1995comparing} test, which tests forecasts of model $A$ against forecasts of model $B$. In the following paper, we compute the multivariate version of the DM test as in \citet{ziel2018day}. The multivariate DM test results in only one statistic for each model that is computed based on the $S$-dimensional vector of losses for each day. Therefore, denote $L_d^A = (L^A_{d,1}, L^A_{d,2}, \dots, L^A_{d,S})'$ and $L^B_{d} = (L^B_{d,1}, L^B_{d,2}, \dots, L^B_{d,S})'$ the vectors of the out-of-sample losses for day $d$ of the models $A$ and $B$, respectively. By $L^Z_{d,s}$  we mean the CRPS$_s$ loss for day $d$ of model $Z$, formally we choose
\begin{align}
L^Z_{d,s} = \frac{1}{R}\sum_{\tau\in r} \text{EVAL}_{(0,\tau,1),(0,\tau,1)}^{Z,d,s}([T_1, T_2))
\end{align}
	with $[T_1, T_2) = [-3.25,-0.5)$.
	In the DM test we consider only the CRPS loss as it is the most important measure in our study.
	The multivariate loss differential series 
	\begin{equation}
	\Delta^{A,B}_{d} = ||L^A_{d}||_q  - ||L^B_{d}||_q  
	\end{equation}
	defines the difference of losses in $||\cdot||_q$ norm, i.e. $||L^{A}_d||_q = \left(\sum_{s=1}^{S} |L^A_{d,s}|^q \right)^{1/q} $, where $q \in \{1,2\}$ in our case. For each model pair, we compute the $p$-value of two one-sided DM tests. The first one is with the null hypothesis $\mathcal{H}_0: \mathbb{E}(\Delta_{d}^{A,B}) \le 0$, i.e. the outperformance of the forecasts of model B by the forecasts of model A. The second test is with the reverse null hypothesis $\mathcal{H}_0^R: \mathbb{E}(\Delta_{d}^{A,B}) \ge 0$, i.e. the outperformance of the forecasts of model A by those of model B. Let us note that these tests are complementary, and we perform them using two norms -- $||\cdot||_1$ and $||\cdot||_2$. Naturally, we assume that the loss differential series is covariance stationary.
	
\section{Results}

\begin{table}
	\centering
	\begingroup\footnotesize
\begin{tabular}{rllll}
	\hline
	& Bias & MAE & RMSE & CRPS \\ 
  \hline
\textbf{Exp.Const} & \cellcolor[rgb]{1,0.621,0.5} { 144.113} & \cellcolor[rgb]{1,0.621,0.5} {220.132} & \cellcolor[rgb]{1,0.621,0.5} {155.464} & \cellcolor[rgb]{1,0.621,0.5} {100.161} \\ 
\textbf{Exp.Lin} & \cellcolor[rgb]{0.52,0.907,0.5} {   1.703} & \cellcolor[rgb]{0.895,1,0.5} {151.349} & \cellcolor[rgb]{0.955,1,0.5} {114.710} & \cellcolor[rgb]{1,1,0.5} { 67.504} \\ 
\textbf{Exp.Quadr} & \cellcolor[rgb]{0.975,1,0.5} { -28.784} & \cellcolor[rgb]{0.63,0.943,0.5} {142.488} & \cellcolor[rgb]{0.675,0.958,0.5} {108.482} & \cellcolor[rgb]{0.81,1,0.5} { 63.410} \\ 
\textbf{Exp.Expon} & \cellcolor[rgb]{1,0.968,0.5} { -40.138} & \cellcolor[rgb]{0.502,0.901,0.5} {138.866} & \cellcolor[rgb]{0.509,0.903,0.5} {105.592} & \cellcolor[rgb]{0.7,0.967,0.5} { 61.761} \\ 
\textbf{Gamma.Const.Const} & \cellcolor[rgb]{1,0.581,0.5} { 156.004} & \cellcolor[rgb]{1,0.588,0.5} {225.793} & \cellcolor[rgb]{1,0.591,0.5} {158.639} & \cellcolor[rgb]{1,0.611,0.5} {100.991} \\ 
\textbf{Gamma.Lin.Const} & \cellcolor[rgb]{0.501,0.9,0.5} {  -0.785} & \cellcolor[rgb]{0.886,1,0.5} {150.973} & \cellcolor[rgb]{0.95,1,0.5} {114.573} & \cellcolor[rgb]{0.96,1,0.5} { 66.631} \\ 
\textbf{Gamma.Lin.Lin} & \cellcolor[rgb]{0.913,1,0.5} {  24.151} & \cellcolor[rgb]{1,0.956,0.5} {163.237} & \cellcolor[rgb]{1,0.93,0.5} {123.187} & \cellcolor[rgb]{1,0.948,0.5} { 72.023} \\ 
\textbf{Gamma.Quadr.Const} & \cellcolor[rgb]{0.977,1,0.5} { -28.934} & \cellcolor[rgb]{0.632,0.944,0.5} {142.568} & \cellcolor[rgb]{0.677,0.959,0.5} {108.524} & \cellcolor[rgb]{0.777,0.992,0.5} { 62.857} \\ 
\textbf{Gamma.Quadr.Lin} & \cellcolor[rgb]{0.944,1,0.5} { -26.498} & \cellcolor[rgb]{0.658,0.953,0.5} {143.301} & \cellcolor[rgb]{0.699,0.966,0.5} {108.908} & \cellcolor[rgb]{0.746,0.982,0.5} { 62.413} \\ 
\textbf{Gamma.Quadr.Quadr} & \cellcolor[rgb]{0.982,1,0.5} { -29.339} & \cellcolor[rgb]{0.772,0.991,0.5} {146.516} & \cellcolor[rgb]{0.828,1,0.5} {111.385} & \cellcolor[rgb]{0.838,1,0.5} { 64.006} \\ 
\textbf{Gamma.Quadr.Expon} & \cellcolor[rgb]{0.897,1,0.5} { -22.970} & \cellcolor[rgb]{0.72,0.973,0.5} {145.052} & \cellcolor[rgb]{0.769,0.99,0.5} {110.128} & \cellcolor[rgb]{0.798,0.999,0.5} { 63.170} \\ 
\textbf{Gamma.Expon.Const} & \cellcolor[rgb]{1,0.968,0.5} { -40.136} & \cellcolor[rgb]{0.505,0.902,0.5} {138.952} & \cellcolor[rgb]{0.51,0.903,0.5} {105.607} & \cellcolor[rgb]{0.663,0.954,0.5} { 61.218} \\ 
\textbf{Gamma.Expon.Lin} & \cellcolor[rgb]{1,0.979,0.5} { -36.848} & \cellcolor[rgb]{0.5,0.9,0.5} {\textbf{138.829}} & \cellcolor[rgb]{0.501,0.9,0.5} {\textbf{105.448}} & \cellcolor[rgb]{0.607,0.936,0.5} { 60.427} \\ 
\textbf{Gamma.Expon.Quadr} & \cellcolor[rgb]{1,0.968,0.5} { -40.102} & \cellcolor[rgb]{0.53,0.91,0.5} {139.659} & \cellcolor[rgb]{0.53,0.91,0.5} {105.957} & \cellcolor[rgb]{0.631,0.944,0.5} { 60.770} \\ 
\textbf{Gamma.Expon.Expon} & \cellcolor[rgb]{1,0.962,0.5} { -41.907} & \cellcolor[rgb]{0.533,0.911,0.5} {139.741} & \cellcolor[rgb]{0.538,0.913,0.5} {106.098} & \cellcolor[rgb]{0.637,0.946,0.5} { 60.845} \\ 
\textbf{GenGam.Const.Const} & \cellcolor[rgb]{1,0.592,0.5} { 152.901} & \cellcolor[rgb]{1,0.5,0.55} {252.987} & \cellcolor[rgb]{1,0.539,0.5} {164.072} & \cellcolor[rgb]{1,0.734,0.5} { 90.424} \\ 
\textbf{GenGam.Lin.Const} & \cellcolor[rgb]{1,0.859,0.5} {  72.961} & \cellcolor[rgb]{1,0.802,0.5} {189.489} & \cellcolor[rgb]{1,0.869,0.5} {129.545} & \cellcolor[rgb]{1,0.989,0.5} { 68.444} \\ 
\textbf{GenGam.Lin.Lin} & \cellcolor[rgb]{0.819,1,0.5} { -17.116} & \cellcolor[rgb]{1,0.971,0.5} {160.708} & \cellcolor[rgb]{1,0.977,0.5} {118.308} & \cellcolor[rgb]{0.592,0.931,0.5} { 60.203} \\ 
\textbf{GenGam.Quadr.Const} & \cellcolor[rgb]{1,0.878,0.5} {  67.173} & \cellcolor[rgb]{1,0.808,0.5} {188.348} & \cellcolor[rgb]{1,0.868,0.5} {129.674} & \cellcolor[rgb]{1,0.987,0.5} { 68.624} \\ 
\textbf{GenGam.Quadr.Lin} & \cellcolor[rgb]{1,0.929,0.5} { -52.039} & \cellcolor[rgb]{1,0.968,0.5} {161.245} & \cellcolor[rgb]{1,0.929,0.5} {123.312} & \cellcolor[rgb]{0.745,0.982,0.5} { 62.396} \\ 
\textbf{GenGam.Quadr.Quadr} & \cellcolor[rgb]{1,0.893,0.5} { -62.730} & \cellcolor[rgb]{1,0.945,0.5} {165.170} & \cellcolor[rgb]{1,0.898,0.5} {126.515} & \cellcolor[rgb]{0.864,1,0.5} { 64.582} \\ 
\textbf{GenGam.Quadr.Expon} & \cellcolor[rgb]{1,0.889,0.5} { -63.908} & \cellcolor[rgb]{0.944,1,0.5} {153.412} & \cellcolor[rgb]{1,0.976,0.5} {118.432} & \cellcolor[rgb]{0.501,0.9,0.5} {\textbf{58.893}} \\ 
\textbf{GenGam.Expon.Const} & \cellcolor[rgb]{1,0.871,0.5} {  69.276} & \cellcolor[rgb]{1,0.794,0.5} {190.731} & \cellcolor[rgb]{1,0.863,0.5} {130.242} & \cellcolor[rgb]{1,0.983,0.5} { 69.003} \\ 
\textbf{GenGam.Expon.Lin} & \cellcolor[rgb]{0.807,1,0.5} {  16.182} & \cellcolor[rgb]{1,0.89,0.5} {174.413} & \cellcolor[rgb]{1,0.917,0.5} {124.556} & \cellcolor[rgb]{0.847,1,0.5} { 64.196} \\ 
\textbf{GenGam.Expon.Quadr} & \cellcolor[rgb]{1,0.905,0.5} { -59.219} & \cellcolor[rgb]{1,0.997,0.5} {156.238} & \cellcolor[rgb]{1,0.956,0.5} {120.499} & \cellcolor[rgb]{0.558,0.919,0.5} { 59.710} \\ 
\textbf{GenGam.Expon.Expon} & \cellcolor[rgb]{0.5,0.9,0.5} {\textbf{-0.735}} & \cellcolor[rgb]{1,0.893,0.5} {173.996} & \cellcolor[rgb]{1,0.904,0.5} {125.897} & \cellcolor[rgb]{0.862,1,0.5} { 64.519} \\ 
\textbf{GenF.Const.Const} & \cellcolor[rgb]{1,0.845,0.5} {  77.112} & \cellcolor[rgb]{1,0.644,0.5} {216.268} & \cellcolor[rgb]{1,0.704,0.5} {146.789} & \cellcolor[rgb]{1,0.905,0.5} { 75.662} \\ 
\textbf{GenF.Lin.Const} & \cellcolor[rgb]{0.654,0.951,0.5} {   8.387} & \cellcolor[rgb]{1,0.912,0.5} {170.659} & \cellcolor[rgb]{1,0.922,0.5} {124.043} & \cellcolor[rgb]{0.686,0.962,0.5} { 61.562} \\ 
\textbf{GenF.Lin.Lin} & \cellcolor[rgb]{1,0.921,0.5} { -54.312} & \cellcolor[rgb]{1,0.955,0.5} {163.464} & \cellcolor[rgb]{1,0.897,0.5} {126.611} & \cellcolor[rgb]{0.697,0.966,0.5} { 61.717} \\ 
\textbf{GenF.Quadr.Const} & \cellcolor[rgb]{0.581,0.927,0.5} {   4.748} & \cellcolor[rgb]{1,0.913,0.5} {170.497} & \cellcolor[rgb]{1,0.919,0.5} {124.345} & \cellcolor[rgb]{0.703,0.968,0.5} { 61.797} \\ 
\textbf{GenF.Quadr.Lin} & \cellcolor[rgb]{1,0.765,0.5} {-101.139} & \cellcolor[rgb]{1,0.932,0.5} {167.424} & \cellcolor[rgb]{1,0.824,0.5} {134.291} & \cellcolor[rgb]{0.873,1,0.5} { 64.764} \\ 
\textbf{GenF.Quadr.Quadr} & \cellcolor[rgb]{1,0.761,0.5} {-102.141} & \cellcolor[rgb]{1,0.901,0.5} {172.586} & \cellcolor[rgb]{1,0.792,0.5} {137.630} & \cellcolor[rgb]{1,0.992,0.5} { 68.198} \\ 
\textbf{GenF.Quadr.Expon} & \cellcolor[rgb]{1,0.736,0.5} {-109.600} & \cellcolor[rgb]{1,0.951,0.5} {164.070} & \cellcolor[rgb]{1,0.837,0.5} {132.864} & \cellcolor[rgb]{0.821,1,0.5} { 63.636} \\ 
\textbf{GenF.Expon.Const} & \cellcolor[rgb]{0.658,0.953,0.5} {   8.630} & \cellcolor[rgb]{1,0.897,0.5} {173.305} & \cellcolor[rgb]{1,0.906,0.5} {125.668} & \cellcolor[rgb]{0.753,0.984,0.5} { 62.511} \\ 
\textbf{GenF.Expon.Lin} & \cellcolor[rgb]{1,0.969,0.5} { -39.876} & \cellcolor[rgb]{1,0.925,0.5} {168.606} & \cellcolor[rgb]{1,0.88,0.5} {128.455} & \cellcolor[rgb]{0.762,0.987,0.5} { 62.641} \\ 
\textbf{GenF.Expon.Quadr} & \cellcolor[rgb]{1,0.745,0.5} {-106.988} & \cellcolor[rgb]{1,0.947,0.5} {164.796} & \cellcolor[rgb]{1,0.827,0.5} {133.924} & \cellcolor[rgb]{0.83,1,0.5} { 63.842} \\ 
\textbf{GenF.Expon.Expon} & \cellcolor[rgb]{1,0.965,0.5} { -41.044} & \cellcolor[rgb]{1,0.903,0.5} {172.291} & \cellcolor[rgb]{1,0.86,0.5} {130.456} & \cellcolor[rgb]{0.828,1,0.5} { 63.797} \\ 
\hline
\end{tabular}
\endgroup
	\caption{Bias, MAE, RMSE and CRPS for the considered models on the 
interval $[T_1,T_2) = [-3.25,-0.5)$. Bolded values indicate the lowest value in each column, instead of bias, where it indicates the value closest to 0.} 
	\label{tab:results}
\end{table}

Table \ref{tab:results} presents the Bias,  MAE, RMSE and CRPS measures (see equation\eqref{eq_crit}
) of the considered models on the 
interval $[T_1,T_2) = [-3.25,-0.5)$ which was used for estimation.
In the table, we observe that the lowest MAE and RMSE are obtained for model \textbf{Gamma.Expon.Lin}. At the same time we see that the values for the other models with gamma distribution and exponential rate function have similar performance, as well as model \textbf{Exp.Expon}, which appears to be the second best in terms of MAE and RMSE. These values indicate that most likely the difference in the performance of modelling the median and mean of models \textbf{Gamma.Expon.Lin} and  \textbf{Exp.Expon} is not statistically significant. The models with generalized gamma and generalized F distributions clearly have difficulties in modelling the central parts of the distribution, but they handle the probabilistic forecasting well. The best performing model in terms of CRPS is \textbf{GenGam.Quadr.Expon}. The CRPS values of other models are mostly satisfying despite the models with constant parameters. These give the worst forecasts in terms of all considered measures.

On the other hand, the values of bias are interesting. We see there that the best models underestimate the true counting process. The reason for such an underestimation may be the fact that the German Intraday Continuous is constantly developing and the number of trades is growing every day. This suggests that there is still some space for improvement of the errors. In the remaining part of this section, we consider only selected best performing models: 2 best models per distribution and from each of the distributions, we choose also the best model with exponential rate function.

	\begin{figure*}[b!]
	\centering
	\subfloat[]{
		\includegraphics[width = 0.475\linewidth]{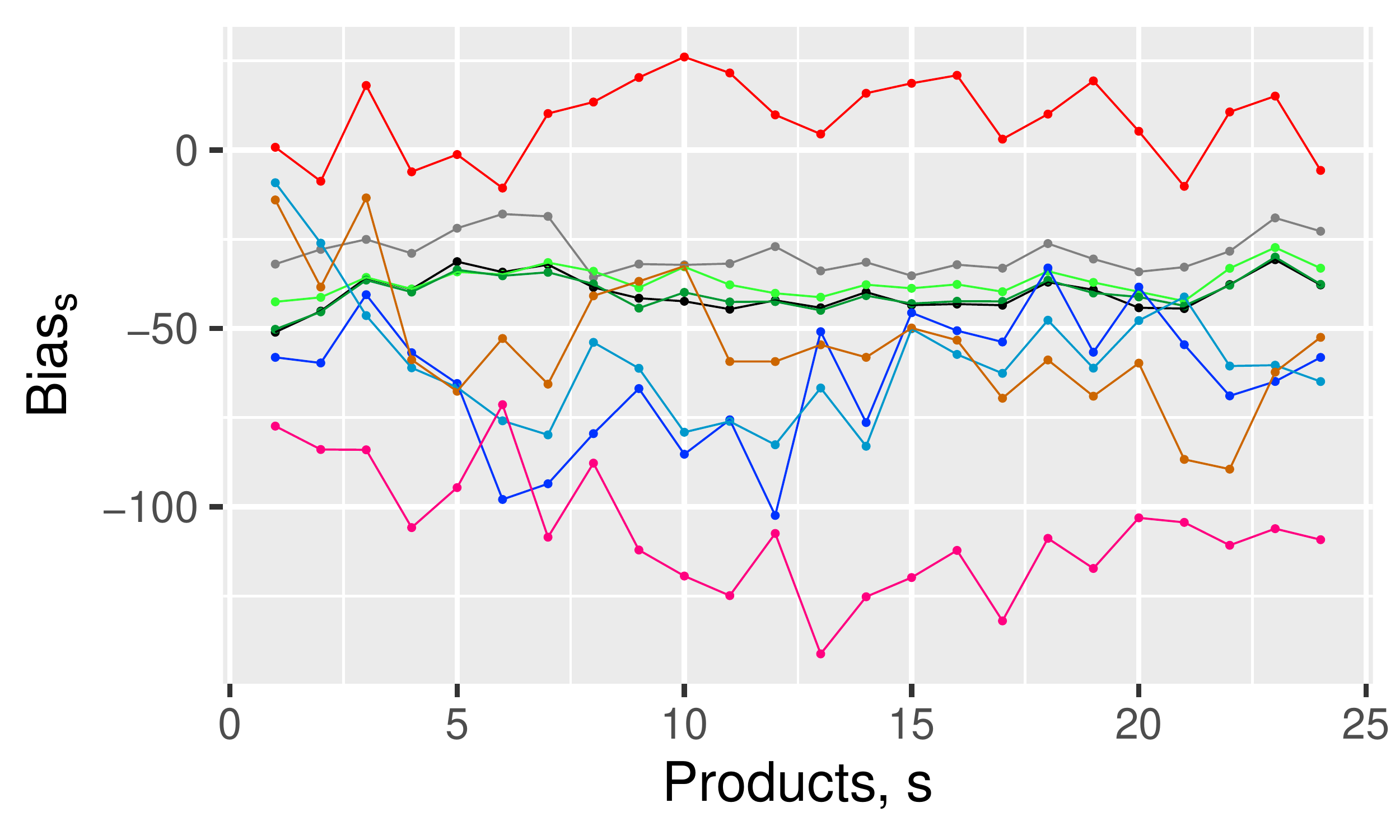}
		\label{fig:biasoverprod}
	}
	~ 
	\subfloat[]{
		\includegraphics[width = 0.475\linewidth]{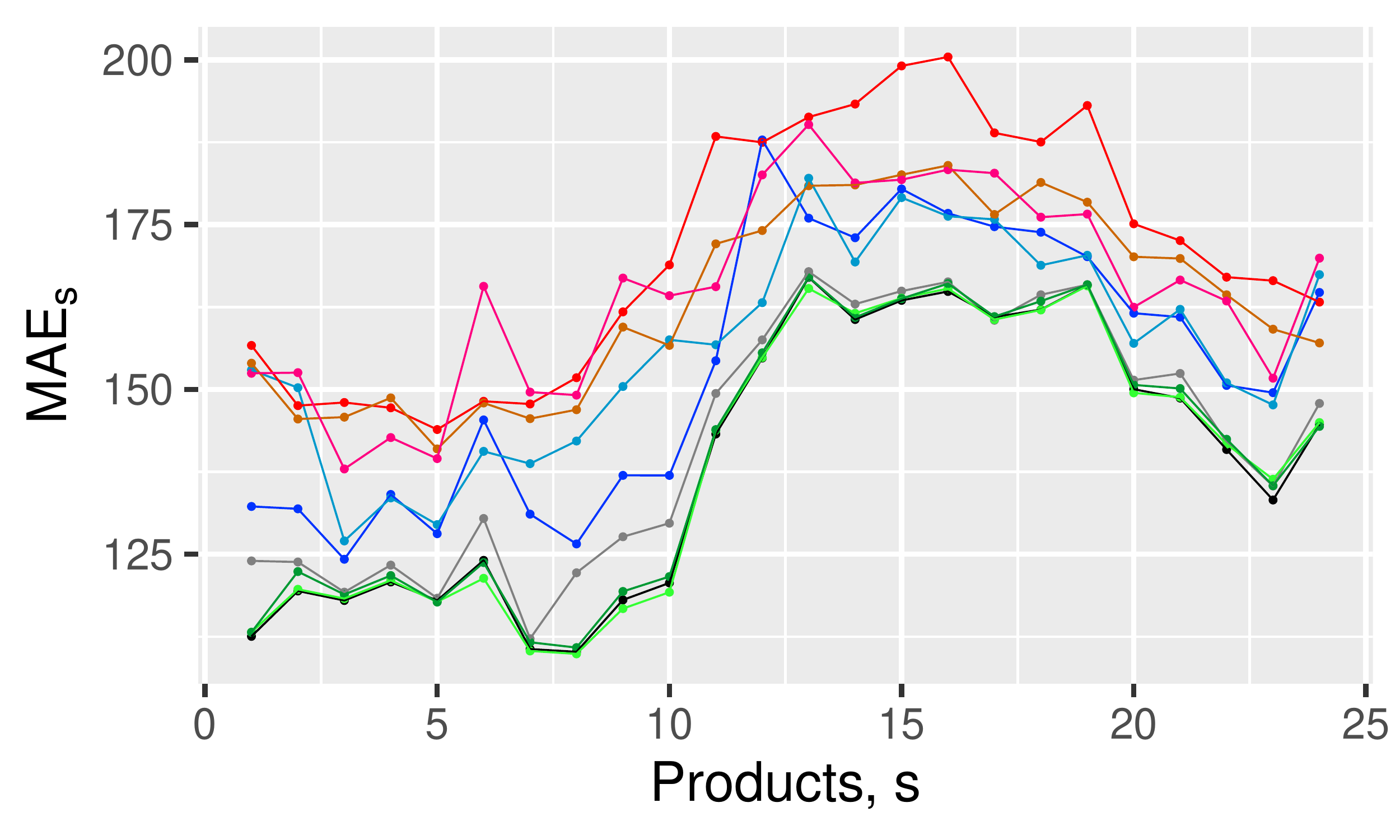}
		\label{fig:l1overprod}
	}\\
	\subfloat[]{
		\includegraphics[width = 0.475\linewidth]{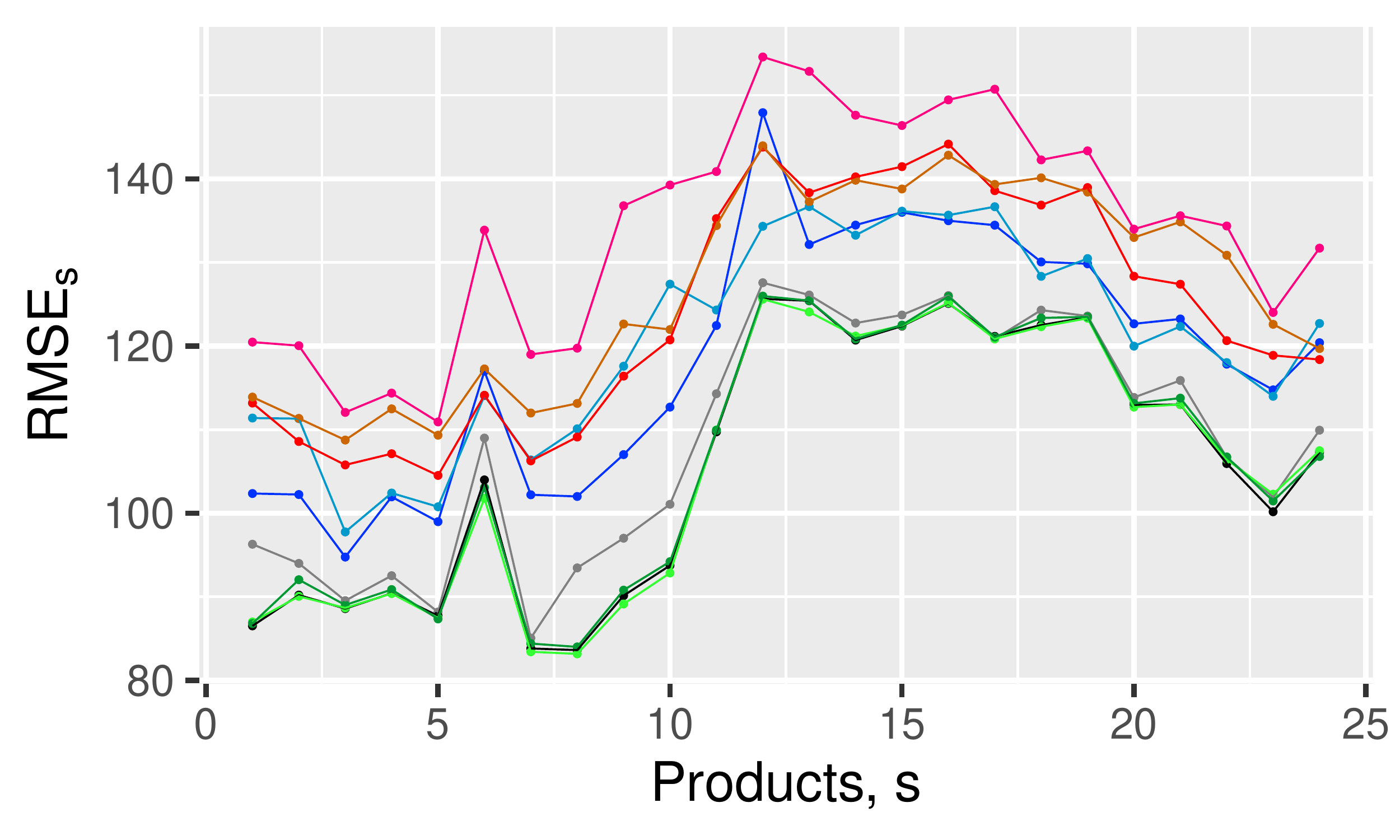}
		\label{fig:l2overprod}
	}
	~
	\subfloat[]{
	\includegraphics[width = 0.475\linewidth]{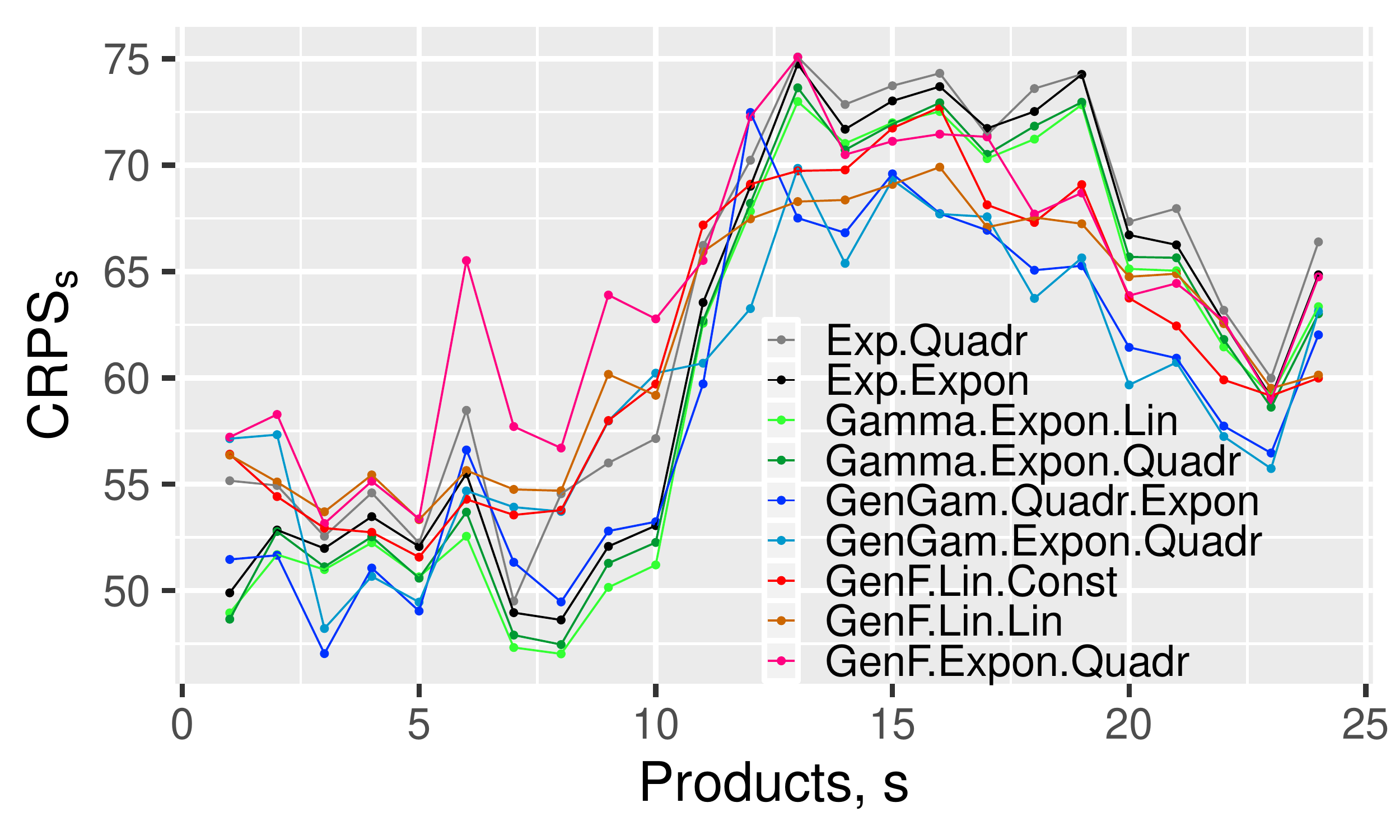}
	\label{fig:crpsoverprod}
	}
	\caption{(a) Bias$_{s}$, (b) MAE$_{s}$, (c) RMSE$_{s}$  and (d) CRPS$_{s}$ of selected best performing models evaluated on $[T_1,T_2) = [-3.25,-0.5)$.}
	\label{fig:errorsoverprods}
\end{figure*}

	\begin{figure*}[b!]
	\centering
	\subfloat[]{
		\includegraphics[width = 0.475\linewidth]{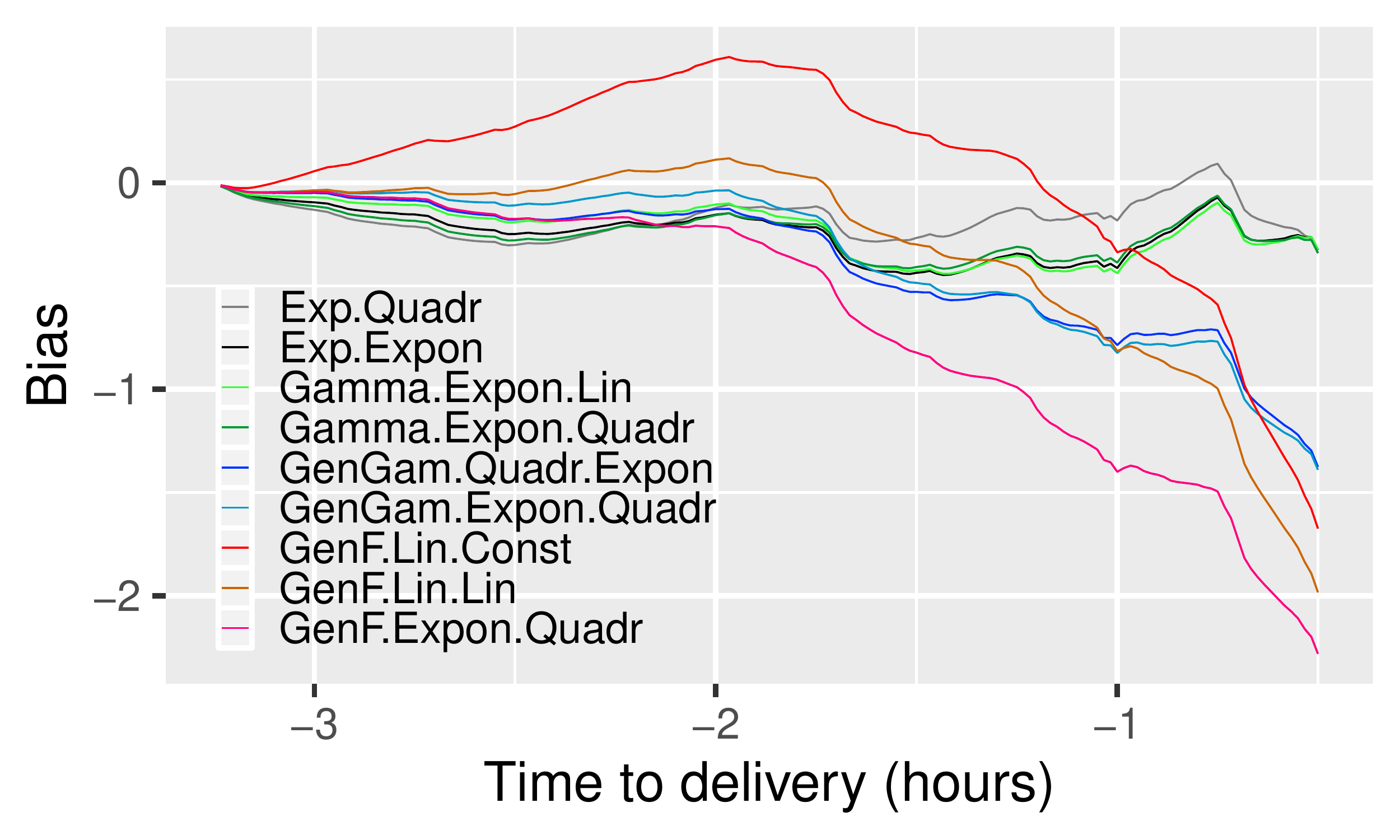}
		\label{fig:biasovertime}
	}
	~
	\subfloat[]{
		\includegraphics[width = 0.475\linewidth]{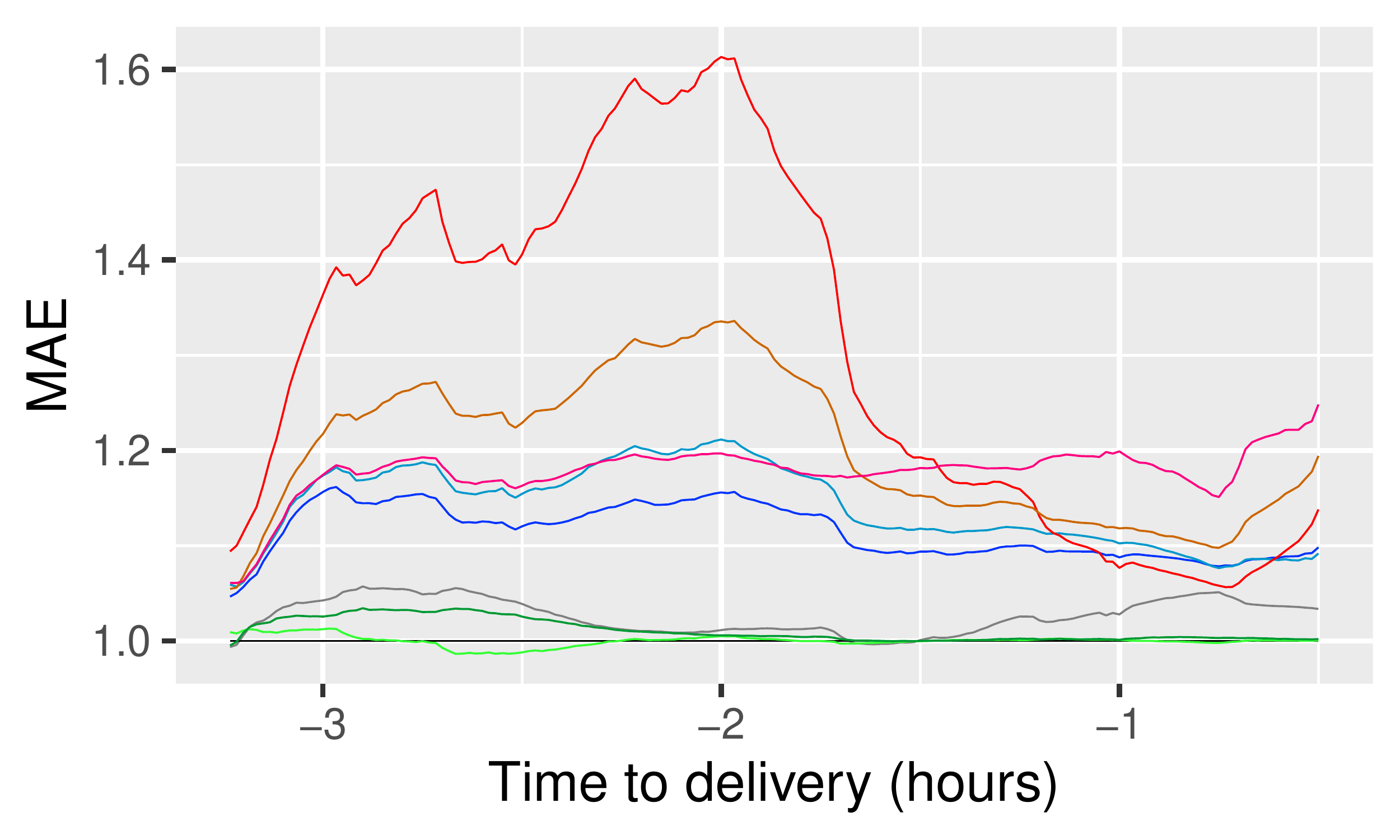}
		\label{fig:l1overtime}
	}\\
	~
	\subfloat[]{
		\includegraphics[width = 0.475\linewidth]{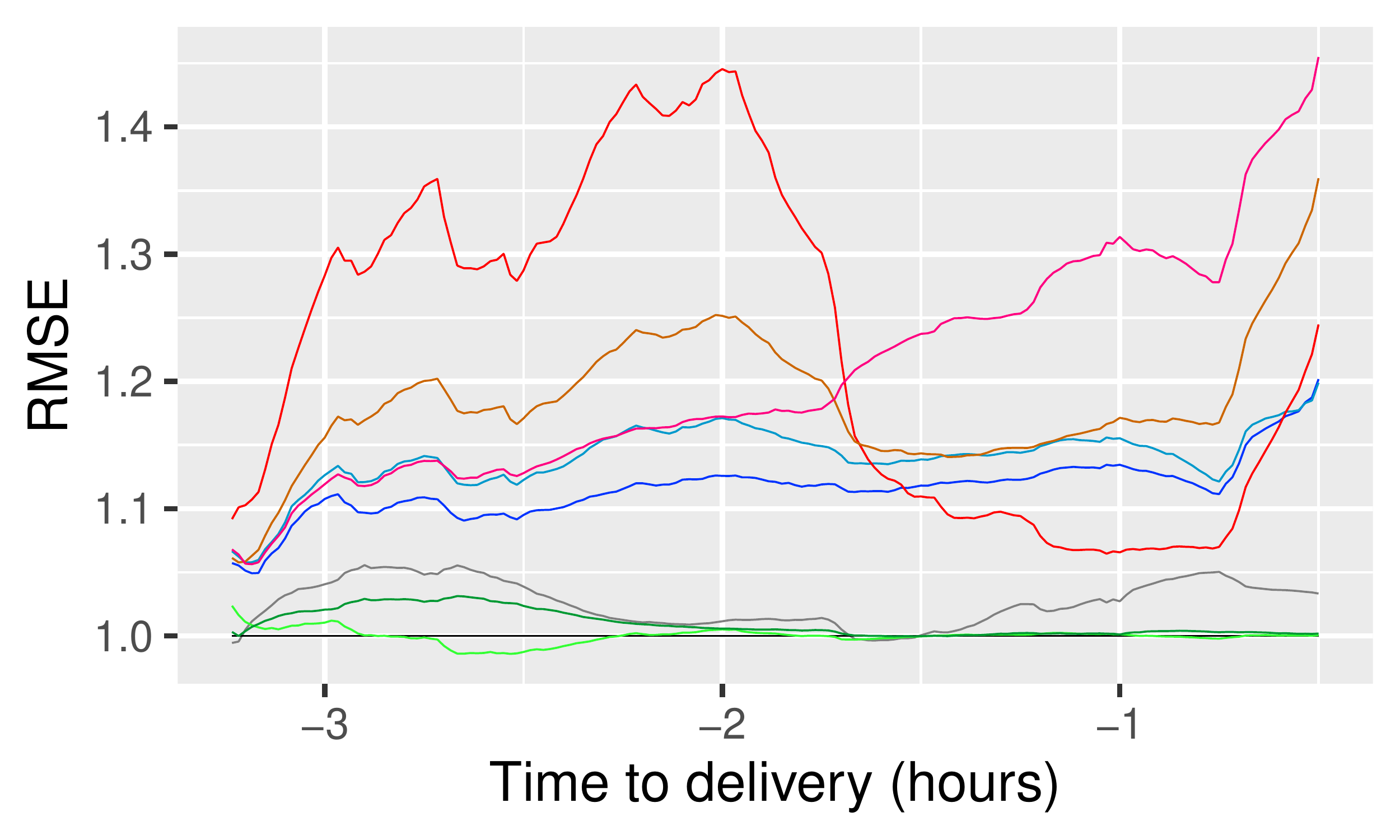}
		\label{fig:l2overtime}
	}
	\subfloat[]{
		\includegraphics[width = 0.475\linewidth]{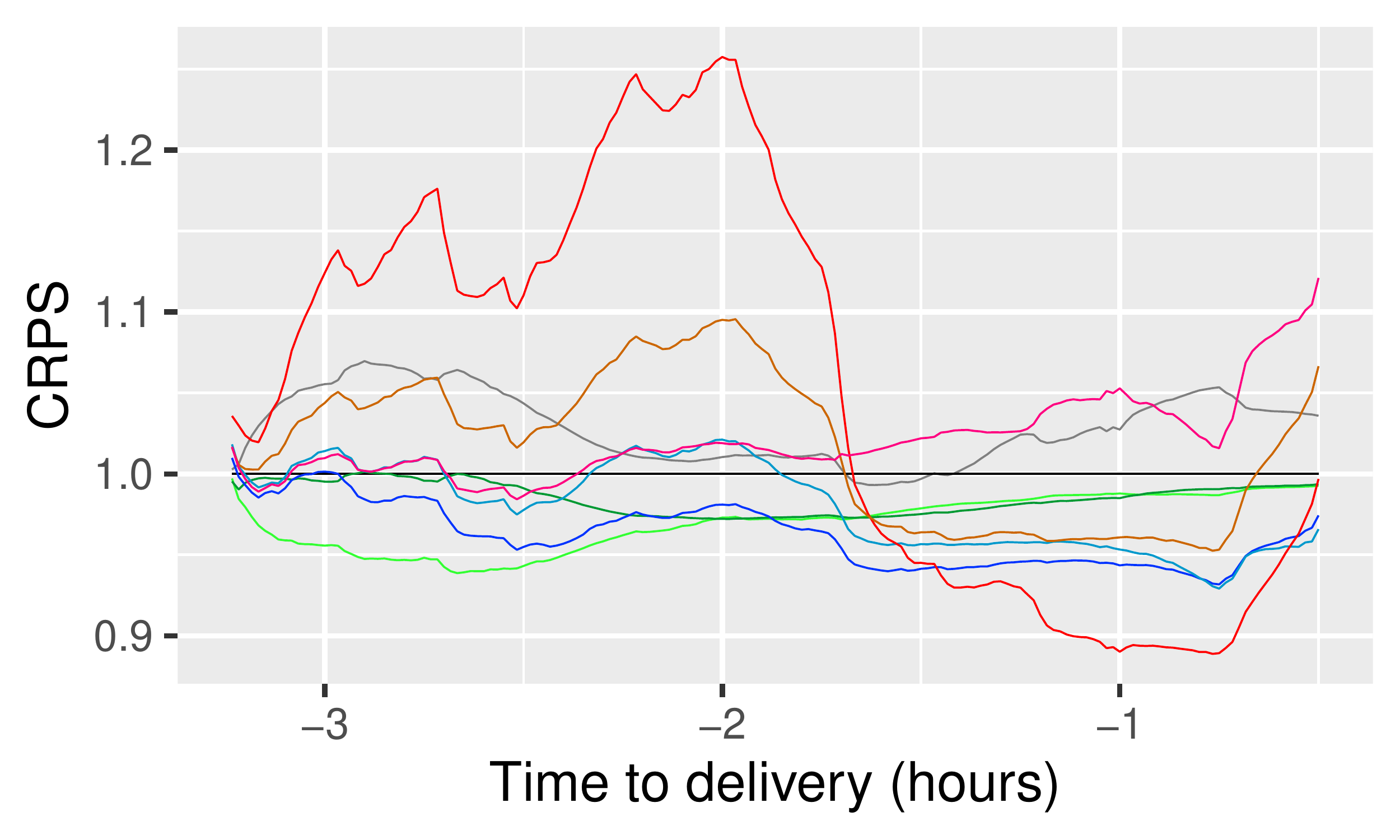}
		\label{fig:crpsovertime}
	}\\
	\subfloat[]{
	\includegraphics[width = 1\linewidth]{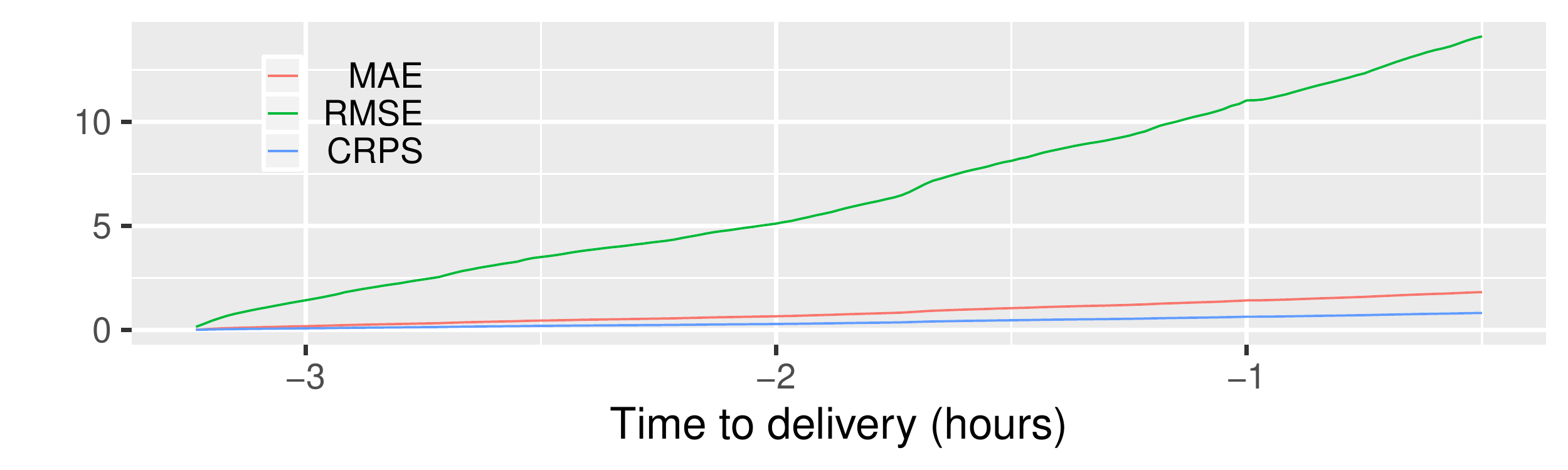}
	\label{fig:measuresovertime}
	}
	\caption{(a) Bias of the selected models, (b) MAE, (c) RMSE, (d) CRPS measures of selected best performing models in relation to the values of the model \textbf{Exp.Expon} and (e)~errors of \textbf{Exp.Expon}  over time to delivery.}
	\label{fig:errorsovertime}
\end{figure*}
Figure \ref{fig:errorsoverprods} contains four graphs that present the performance of the selected models' forecasts over products 
$s\in \{1,\ldots, S\}$. In Figure \ref{fig:biasoverprod}, we observe that all the models despite the \textbf{GenF.Lin.Const} underestimate the true trajectory for all products. There is no clear pattern of the underestimation, but we see that the bias of best models in terms of MAE and RMSE is almost constant over products, whereas the bias of the best model in terms of CRPS varies significantly. 
Figures \ref{fig:l1overprod}, \ref{fig:l2overprod} and \ref{fig:crpsoverprod} show that all models handle the simulation of night hours' transactions better than the simulation of the day hours'. An interesting spike in both measures can be observed for $s=6$. The reason for that might be some outliers as the spike is higher in terms of RMSE than in terms of MAE or CRPS. Furthermore, we see that the models with exponential and gamma distribution are clearly better than the others across all products in terms of MAE and RMSE. It is different for CRPS, where for most hours the best models are the GenGam models, but between hours 6 and 10 the gamma distributed models appear to be better.

\begin{figure}[p!tb]
	\centering
	\includegraphics[width = 1\linewidth]{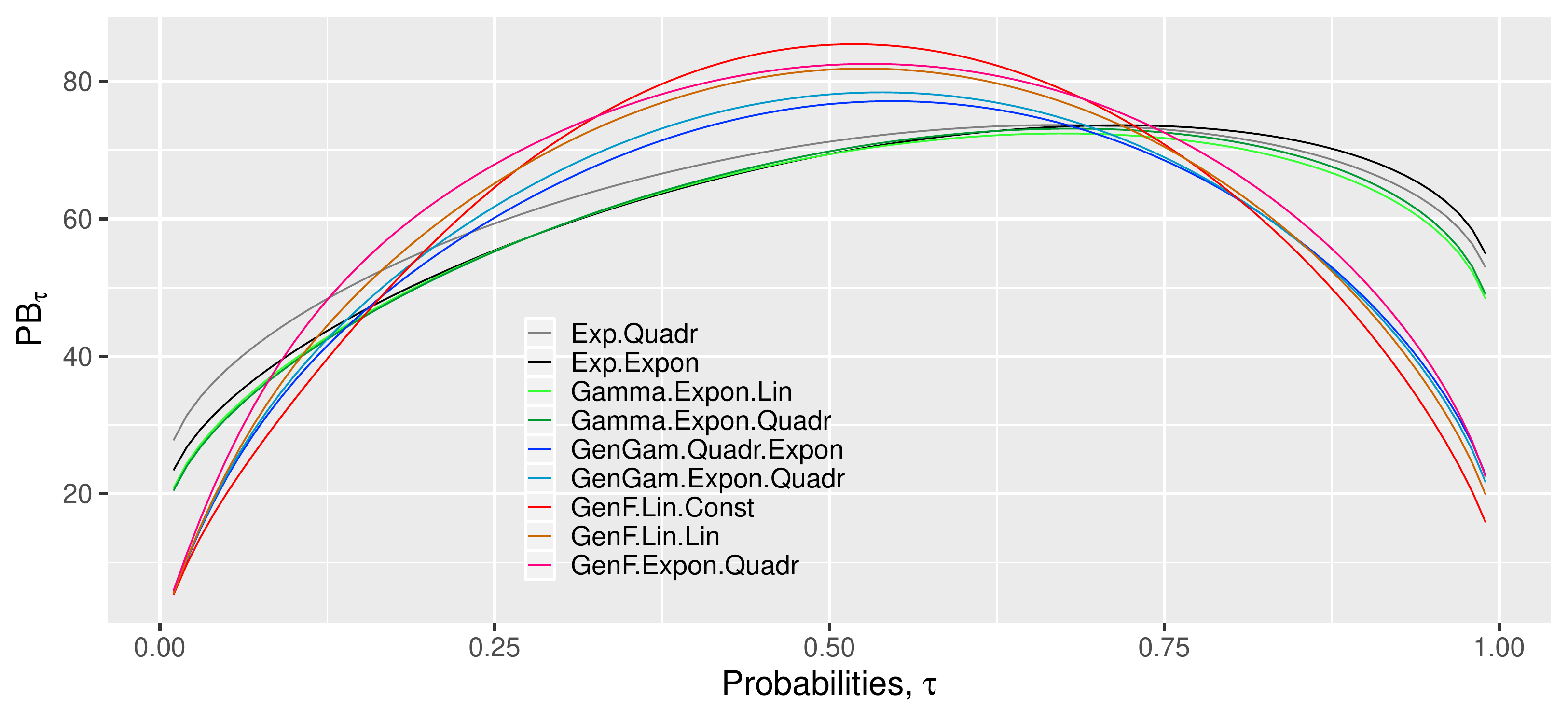}
	\caption{PB$_{\tau}$ of selected best performing models over probabilities 
	$\tau \in \{0.01,0.02,\ldots,0.99\}$.}
	\label{fig:crpsvsprobs}
\end{figure}

Figure \ref{fig:errorsovertime} is analogous to Figure \ref{fig:errorsoverprods}, but the measures are calculated over the time to delivery. This means that we calculate the values (see equation \eqref{eq_crit}) on many short time intervals, i.e. we apply a minute grid of the time-frame $[-3.25,-0.5)$. This way we can understand which part of the trajectory is forecasted the most and the least precisely. Figure \ref{fig:biasovertime} shows that the most underestimated is the last 1.5 hour of the counting process, but the models with gamma and exponential distribution handle this period better than the GenGam and GenF models. 
Figures \ref{fig:l1overtime} and \ref{fig:l2overtime} show again the Gamma and Exp models with exponential rate function give mostly the lowest MAE and RMSE. In Figure \ref{fig:crpsovertime}, we see that \textbf{Gamma.Expon.Lin} model has lowest CRPS in the first half of the forecasting horizon and, surprisingly, in the second half the lowest CRPS value is obtained for model \textbf{GenF.Lin.Const}. The performance of \textbf{GenGam.Quadr.Expon} is mostly stable over the time to delivery. Figure \ref{fig:crpsvsprobs} shows the pinball loss results of the models over probability values. Based on it, it is clear that the Gamma and Exp models forecast better the central quantiles, i.e. between 0.2 and 0.7, whereas the others are better forecasted by GenGam and GenF models. The figure indicates that the overall CRPS performance can be significantly improved.

	\begin{figure*}[p!tb]
	\centering
	\subfloat[]{
		\includegraphics[width = 0.475\linewidth]{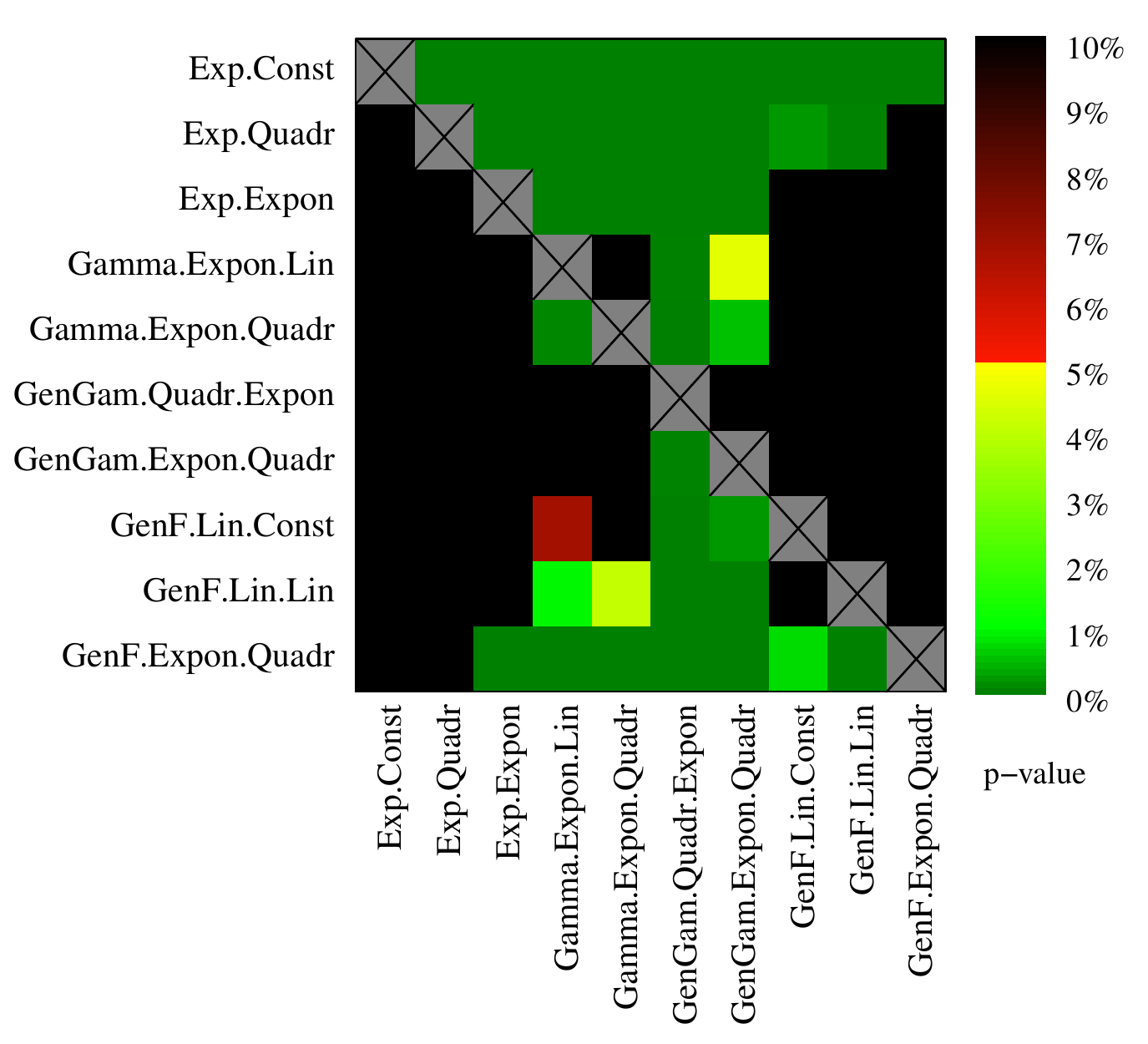}
		\label{fig:dmtest_MAE_l1}
	}%
	~ 
	\subfloat[]{
		\includegraphics[width = 0.475\linewidth]{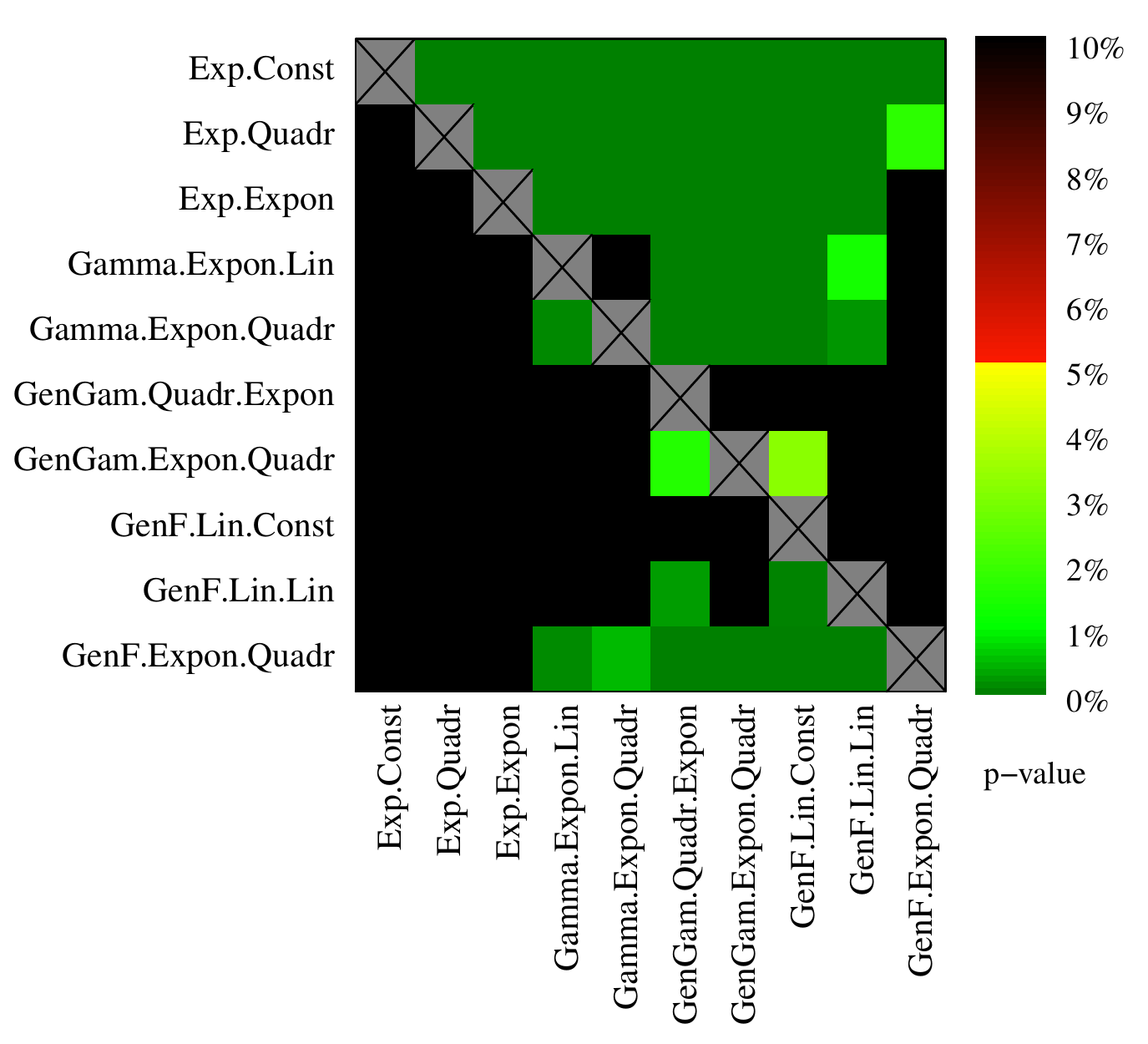}
		\label{fig:dmtest_MAE_l2}
	}
	\caption{Results of the Diebold-Mariano test. (a) presents the $p$-values for the $||\cdot||_1$ norm with the CRPS loss, (b) the values for the $||\cdot||_2$ norm with the CRPS loss. The figures use a heat map to indicate the range of the $p$-values. The closer they are to zero ($\to$ dark green), the more significant the difference is between forecasts of X-axis model (better) and forecasts of the Y-axis model (worse).}
	\label{fig:DMtest}

\end{figure*}

To draw statistically significant conclusions, we perform a Diebold-Mariano test. The results are presented in Figure \ref{fig:DMtest}. Let us remind that we use the CRPS as the loss series as we believe that it is the most important measure in this study. Based on it, it is clear that the \textbf{GenGam.Quadr.Expon} model is significantly the best. Interestingly, the test in norm $||\cdot||_2$ indicates a very good performance of the GenF models. Although, in this norm none of the other models is significantly better than the \textbf{GenGam.Quadr.Expon}.

	\begin{figure*}[b!]
	\centering
	\subfloat[]{
		\includegraphics[width = 1\linewidth]{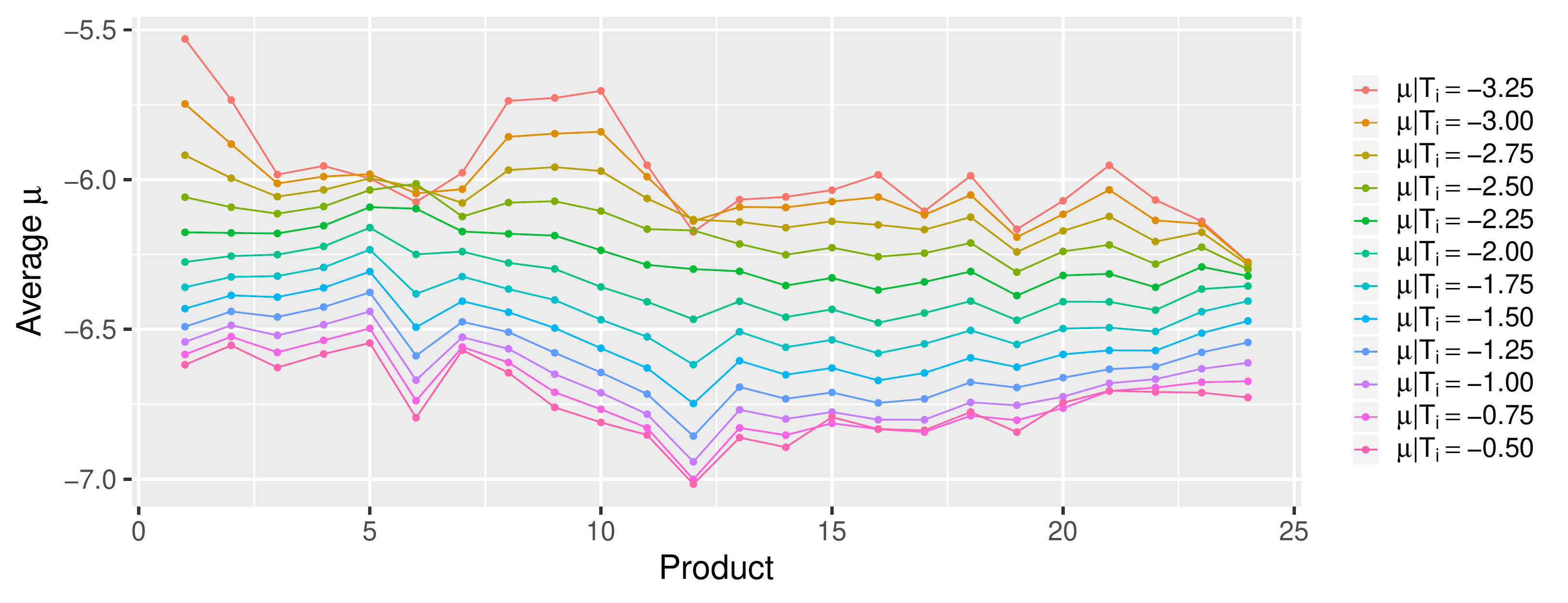}
		\label{fig:coefsoverprod}
	}\\ 
	\subfloat[]{
		\includegraphics[width = 1\linewidth]{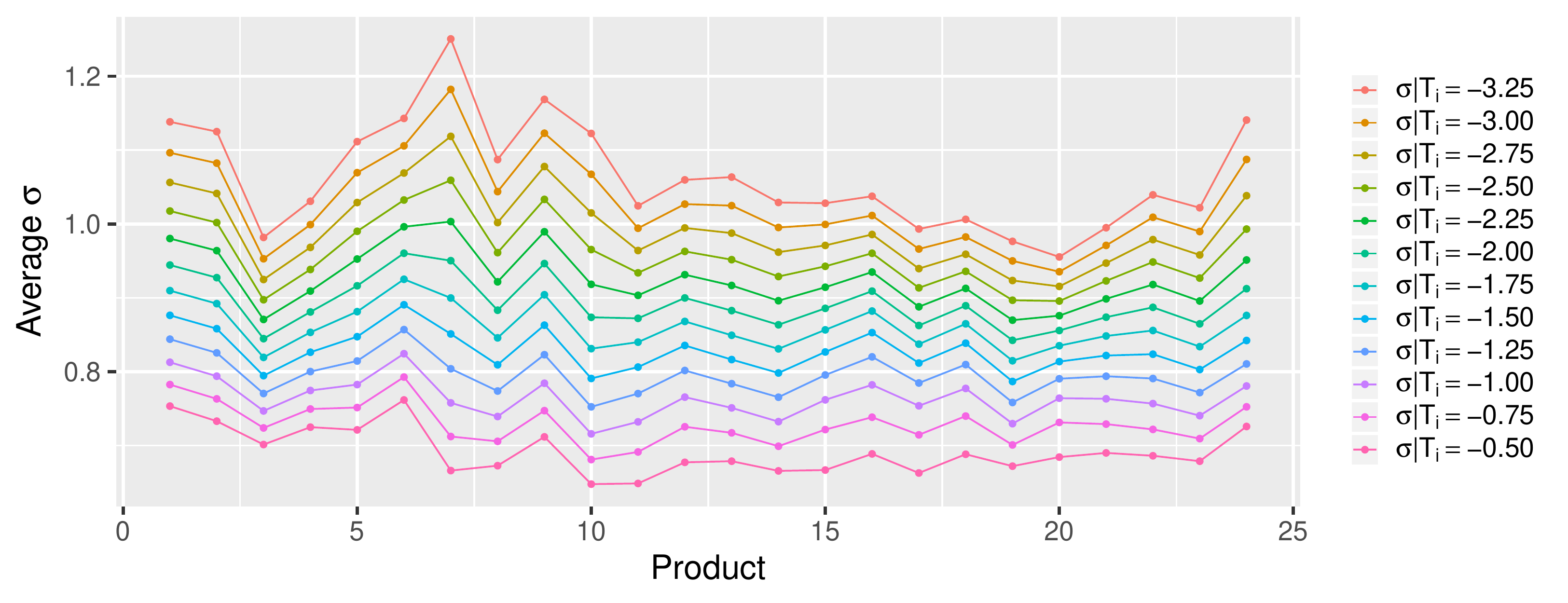}
		\label{fig:coefsoveriter}
	}
	\caption{Average (a) location $\mu$ and (b) scale $\sigma$ parameters of model \textbf{GenGam.Quadr.Expon} over products at time $T_i \in \{ -3.25, -3, \dots, -0.5\}$}
	\label{fig:coefs}
\end{figure*}

Let us now take a look at the values of the time-varying coefficients of the model \textbf{GenGam.Quadr.Expon}. Let us recall that in this model we estimate the inter-arrival time $X_i^{d,s}$ assuming generalized gamma distribution with the 
time-varying coefficients $\mu(\theta,t) = -\log(\beta(\theta_2,t)/\alpha(\theta_1,t))$ and $\sigma(\theta,t) = 1/\sqrt{\alpha(\theta_1,t)}$ and constant $Q$,  where $\beta(\theta_2,t)$ and $\alpha(\theta_1,t)$ are the rate and shape parameters of the standard gamma distribution. In this model $\beta(\theta_2,t)$ is a quadratic function and $\alpha(\theta_1,t)$ is an exponential function.
In Figure \ref{fig:coefs}, we analyse the behaviour of $\mu$ and $\sigma$  parameters over products at different time. 
Figure~\ref{fig:coefsoverprod} shows that the values of $\mu$ are similar over products, but differ significantly over time. That is to say, the closer we are to the gate closure, the lower the location coefficient, which means that the transactions appear more often. Figure \ref{fig:coefsoveriter} shows a very similar behaviour for the scale parameter $\sigma$. In this case the closer we are to the gate closure, the less vary the inter-arrival times. 

\section{Conclusion}
We described the novel problem of estimation and simulation of the transaction arrival process in intraday electricity markets. The approach is not complicated and the presented methods are easy to implement. The paper fills the gap in the literature regarding the estimation and simulation of the transaction arrival process in the intraday electricity markets and thus is a major contribution to this field of research. The outcome of the conducted study is very satisfying.

Using the aforementioned approach, we utilized an exemplary rolling window forecasting study based on the German Intraday Continuous market. We assumed four probability distributions of the inter-arrival times: exponential, gamma, generalized gamma and generalized F distributions. We performed the maximum likelihood estimation of the distributions, assuming time-dependence of some of their coefficients. Then, using the estimated distributions we simulated new trajectories which we evaluated using the functional bias, MAE, RMSE and CRPS. 

The results showed that the forecasting error can be significantly reduced, comparing to the most standard benchmark, which was the homogeneous Poisson process. The best in terms of forecasting of the central part of the distribution of the transaction arrivals, i.e. the mean or median are the exponential and gamma distributions with exponential rate function. In terms of the more meaningful and thus more important CRPS significantly the best forecasts were obtained from the generalized gamma model with quadratic rate function and exponential shape function. 
	
This field of research can be easily developed further. A possible direction is considering other probability distributions of the inter-arrival times. Another possibility would be using more complex distributions' parameter functions, e.g. Hawkes process-like, which is widely applied to modelling the transaction time arrivals in the financial markets, see e.g. \citet{hewlett2006clustering} or \citet{bacry2015hawkes}. The parameter functions could be also modelled using smoothing kernel or splines. To avoid overestimation, regularization methods should be considered.
	
\section*{Acknowledgments}
	
	This research article was partially supported by the German Research Foundation (DFG, Germany) and the National Science Center (NCN, Poland) through BEETHOVEN grant no. 2016/23/G/HS4/01005.
	
	\bibliographystyle{chicago}

\bibliography{bibliography}	

\begin{thebibliography}{}

\bibitem[\protect\citeauthoryear{A{\"\i}d, Gruet, and Pham}{A{\"\i}d
  et~al.}{2016}]{aid2016optimal}
A{\"\i}d, R., P.~Gruet, and H.~Pham (2016).
\newblock An optimal trading problem in intraday electricity markets.
\newblock {\em Mathematics and Financial Economics\/}~{\em 10\/}(1), 49--85.

\bibitem[\protect\citeauthoryear{Andrade, Filipe, Reis, and Bessa}{Andrade
  et~al.}{2017}]{andrade2017probabilistic}
Andrade, J.~R., J.~Filipe, M.~Reis, and R.~J. Bessa (2017).
\newblock {Probabilistic Price Forecasting for Day-Ahead and Intraday Markets:
  Beyond the Statistical Model}.
\newblock {\em Sustainability\/}~{\em 9\/}(11), 1990.

\bibitem[\protect\citeauthoryear{Bacry, Mastromatteo, and Muzy}{Bacry
  et~al.}{2015}]{bacry2015hawkes}
Bacry, E., I.~Mastromatteo, and J.-F. Muzy (2015).
\newblock Hawkes processes in finance.
\newblock {\em Market Microstructure and Liquidity\/}~{\em 1\/}(01), 1550005.

\bibitem[\protect\citeauthoryear{Diebold and Mariano}{Diebold and
  Mariano}{1995}]{diebold1995comparing}
Diebold, F. and R.~Mariano (1995).
\newblock {Comparing Predictive Accuracy}.
\newblock {\em Journal of Business \& Economic Statistics\/}~{\em 13\/}(3),
  253--63.

\bibitem[\protect\citeauthoryear{Diebold}{Diebold}{2015}]{diebold2015comparing}
Diebold, F.~X. (2015).
\newblock Comparing predictive accuracy, twenty years later: A personal
  perspective on the use and abuse of diebold--mariano tests.
\newblock {\em Journal of Business \& Economic Statistics\/}~{\em 33\/}(1),
  1--1.

\bibitem[\protect\citeauthoryear{Ghalanos and Theussl}{Ghalanos and
  Theussl}{2015}]{rsolnp}
Ghalanos, A. and S.~Theussl (2015).
\newblock {\em Rsolnp: General Non-linear Optimization Using Augmented Lagrange
  Multiplier Method}.
\newblock R package version 1.16.

\bibitem[\protect\citeauthoryear{Gonz{\'a}lez-Aparicio and
  Zucker}{Gonz{\'a}lez-Aparicio and Zucker}{2015}]{gonzalez2015impact}
Gonz{\'a}lez-Aparicio, I. and A.~Zucker (2015).
\newblock Impact of wind power uncertainty forecasting on the market
  integration of wind energy in spain.
\newblock {\em Applied energy\/}~{\em 159}, 334--349.

\bibitem[\protect\citeauthoryear{Graf~von Luckner, Cartea, Jaimungal, and
  Kiesel}{Graf~von Luckner et~al.}{2017}]{von2017optimal}
Graf~von Luckner, N., {\'A}.~Cartea, S.~Jaimungal, and R.~Kiesel (2017).
\newblock Optimal market maker pricing in the german intraday power market.

\bibitem[\protect\citeauthoryear{Hewlett}{Hewlett}{2006}]{hewlett2006clustering}
Hewlett, P. (2006).
\newblock Clustering of order arrivals, price impact and trade path
  optimisation.
\newblock In {\em Workshop on Financial Modeling with Jump processes, Ecole
  Polytechnique}, pp.\  6--8.

\bibitem[\protect\citeauthoryear{Jackson}{Jackson}{2016}]{jackson2016flexsurv}
Jackson, C.~H. (2016).
\newblock flexsurv: a platform for parametric survival modeling in r.
\newblock {\em Journal of Statistical Software\/}~{\em 70}.

\bibitem[\protect\citeauthoryear{Kath}{Kath}{2019}]{kath2019modeling}
Kath, C. (2019).
\newblock Modeling intraday markets under the new advances of the cross-border
  intraday project (xbid): Evidence from the german intraday market.
\newblock {\em Energies\/}~{\em 12\/}(22), 4339.

\bibitem[\protect\citeauthoryear{Kiesel and Paraschiv}{Kiesel and
  Paraschiv}{2017}]{Kiesel2017}
Kiesel, R. and F.~Paraschiv (2017).
\newblock Econometric analysis of 15-minute intraday electricity prices.
\newblock {\em Energy Economics\/}~{\em 64}, 77--90.

\bibitem[\protect\citeauthoryear{Kulakov and Ziel}{Kulakov and
  Ziel}{2019}]{kulakov2019impact}
Kulakov, S. and F.~Ziel (2019).
\newblock The impact of renewable energy forecasts on intraday electricity
  prices.
\newblock {\em arXiv preprint arXiv:1903.09641\/}.

\bibitem[\protect\citeauthoryear{Monteiro, Ramirez-Rosado, Fernandez-Jimenez,
  and Conde}{Monteiro et~al.}{2016}]{monteiro2016short}
Monteiro, C., I.~J. Ramirez-Rosado, L.~A. Fernandez-Jimenez, and P.~Conde
  (2016).
\newblock {Short-Term Price Forecasting Models Based on Artificial Neural
  Networks for Intraday Sessions in the Iberian Electricity Market}.
\newblock {\em Energies\/}~{\em 9\/}(9), 721.

\bibitem[\protect\citeauthoryear{Narajewski and Ziel}{Narajewski and
  Ziel}{2019}]{narajewski2018econometric}
Narajewski, M. and F.~Ziel (2019).
\newblock Econometric modelling and forecasting of intraday electricity prices.
\newblock {\em Journal of Commodity Markets\/}, 100107.

\bibitem[\protect\citeauthoryear{Nowotarski and Weron}{Nowotarski and
  Weron}{2018}]{nowotarski2018recent}
Nowotarski, J. and R.~Weron (2018).
\newblock Recent advances in electricity price forecasting: A review of
  probabilistic forecasting.
\newblock {\em Renewable and Sustainable Energy Reviews\/}~{\em 81},
  1548--1568.

\bibitem[\protect\citeauthoryear{Pape, Hagemann, and Weber}{Pape
  et~al.}{2016}]{pape2016fundamentals}
Pape, C., S.~Hagemann, and C.~Weber (2016).
\newblock Are fundamentals enough? explaining price variations in the german
  day-ahead and intraday power market.
\newblock {\em Energy Economics\/}~{\em 54}, 376--387.

\bibitem[\protect\citeauthoryear{Prentice}{Prentice}{1974}]{prentice1974log}
Prentice, R.~L. (1974).
\newblock A log gamma model and its maximum likelihood estimation.
\newblock {\em Biometrika\/}~{\em 61\/}(3), 539--544.

\bibitem[\protect\citeauthoryear{Prentice}{Prentice}{1975}]{prentice1975discrimination}
Prentice, R.~L. (1975).
\newblock Discrimination among some parametric models.
\newblock {\em Biometrika\/}~{\em 62\/}(3), 607--614.

\bibitem[\protect\citeauthoryear{Stacy et~al.}{Stacy
  et~al.}{1962}]{stacy1962generalization}
Stacy, E.~W. et~al. (1962).
\newblock A generalization of the gamma distribution.
\newblock {\em The Annals of mathematical statistics\/}~{\em 33\/}(3),
  1187--1192.

\bibitem[\protect\citeauthoryear{Uniejewski, Marcjasz, and Weron}{Uniejewski
  et~al.}{2019}]{uniejewski2018understanding}
Uniejewski, B., G.~Marcjasz, and R.~Weron (2019).
\newblock Understanding intraday electricity markets: Variable selection and
  very short-term price forecasting using lasso.
\newblock {\em International Journal of Forecasting\/}.

\bibitem[\protect\citeauthoryear{Viehmann}{Viehmann}{2017}]{Viehmann2017}
Viehmann, J. (2017, Jun).
\newblock {State of the German Short-Term Power Market}.
\newblock {\em Zeitschrift f{\"u}r Energiewirtschaft\/}~{\em 41\/}(2), 87--103.

\bibitem[\protect\citeauthoryear{Ye}{Ye}{1987}]{rsolnpphd}
Ye, Y. (1987).
\newblock {\em Interior Algorithms for Linear, Quadratic, and Linearly
  Constrained Non-Linear Programming}.
\newblock Ph.\ D. thesis, Department of {ESS}, Stanford University.

\bibitem[\protect\citeauthoryear{Ziel}{Ziel}{2017}]{ziel2017modeling}
Ziel, F. (2017).
\newblock Modeling the impact of wind and solar power forecasting errors on
  intraday electricity prices.
\newblock In {\em 2017 14th International Conference on the European Energy
  Market (EEM)}, pp.\  1--5. IEEE.

\bibitem[\protect\citeauthoryear{Ziel and Weron}{Ziel and
  Weron}{2018}]{ziel2018day}
Ziel, F. and R.~Weron (2018).
\newblock Day-ahead electricity price forecasting with high-dimensional
  structures: Univariate vs. multivariate modeling frameworks.
\newblock {\em Energy Economics\/}~{\em 70}, 396--420.

\end{thebibliography}

\end{document}